\documentclass[default,iicol]{sn-jnl}% Default with double column layout

\usepackage{subfig}
\usepackage{multirow}
\usepackage{booktabs}
\usepackage{multirow}
\usepackage{longtable}
\usepackage{lscape}
\usepackage{color}

\usepackage{float}
 \usepackage{etoolbox}
\makeatletter
\patchcmd{\ps@headings}
{\hbox to \hsize{\hfill Springer Nature 2021 \LaTeX\ template\hfill}}
{\hbox to \hsize{}}
{}
{}
\patchcmd{\ps@titlepage}
{\hbox to \hsize{\hfill Springer Nature 2021 \LaTeX\ template\hfill}}
{\hbox to \hsize{}}
{}
{}
\makeatother
\makeatletter
\patchcmd{\ps@headings}
{\hbox to \hsize{\hfill Springer Nature 2021 \LaTeX\ template\hfill}}
{\hbox to \hsize{}}
{}
{}
\patchcmd{\ps@headings}
{\hbox to \hsize{\hfill Springer Nature 2021 \LaTeX\ template\hfill}}
{\hbox to \hsize{}}
{}
{}
\patchcmd{\ps@titlepage}
{\hbox to \hsize{\hfill Springer Nature 2021 \LaTeX\ template\hfill}}
{\hbox to \hsize{}}
{}
{}
\makeatother
\jyear{2022}%

\theoremstyle{thmstyleone}%
\theoremstyle{thmstyletwo}%

\theoremstyle{thmstylethree}%

\begin{document}

\title[Article Title]{A Review of Intelligent Music Generation Systems}

%%=============================================================%%
%% Prefix	-> \pfx{Dr}
%% GivenName	-> \fnm{Joergen W.}
%% Particle	-> \spfx{van der} -> surname prefix
%% FamilyName	-> \sur{Ploeg}
%% Suffix	-> \sfx{IV}
%% NatureName	-> \tanm{Poet Laureate} -> Title after name
%% Degrees	-> \dgr{MSc, PhD}
%% \author*[1,2]{\pfx{Dr} \fnm{Joergen W.} \spfx{van der} \sur{Ploeg} \sfx{IV} \tanm{Poet Laureate} 
%%                 \dgr{MSc, PhD}}\email{iauthor@gmail.com}
%%=============================================================%%
\author[1]{\fnm{Lei} \sur{Wang}}\email{wanglei@tongji.edu.cn}
\author*[1]{\fnm{Ziyi} \sur{Zhao}}\email{zhaozi1@tongji.edu.cn}
\author[1]{\fnm{Hanwei} \sur{Liu}}\email{liuhw1@tongji.edu.cn}
\author[1]{\fnm{Junwei} \sur{Pang}}\email{pangjw@tongji.edu.cn}
\author[2]{\fnm{Yi} \sur{Qin}}\email{qinyi@shcmusic.edu.cn}
\author[1]{\fnm{Qidi} \sur{Wu}}

\affil*[1]{\orgdiv{College of Electric and Information Engineering}, \orgname{Tongji University}, \orgaddress{\street{Cao'an Street}, \city{Shanghai}, \postcode{201804}, \country{China}}}
 
\affil[2]{\orgdiv{Department of Music Engineering}, \orgname{Shanghai Conservatory of Music}, \orgaddress{\street{Fenyang street}, \city{Shanghai}, \postcode{200031}, \country{China}}}

%%==================================%%
%% sample for unstructured abstract %%
%%==================================%%

\abstract{With the introduction of ChatGPT, the public's perception of AI-generated content (AIGC) has begun to reshape. Artificial intelligence has significantly reduced the barrier to entry for non-professionals in creative endeavors, enhancing the efficiency of content creation. Recent advancements have seen significant improvements in the quality of \textcolor{black}{symbolic music generation}, which is enabled by the use of modern generative algorithms to extract patterns implicit in a piece of music based on rule constraints or a musical corpus. \textcolor{black}{Nevertheless, existing literature reviews tend to present a conventional and conservative perspective on future development trajectories, with a notable absence of thorough benchmarking of generative models.} This paper provides a survey and analysis of recent intelligent music generation techniques, outlining their respective characteristics and discussing existing methods for evaluation. Additionally, the paper compares the different characteristics of music generation techniques in the East and West as well as analysing the field's development prospects.}

\maketitle
\section{Introduction}\label{sec1}

\begin{figure*}[!h]
\centering
\includegraphics[width=1\textwidth]{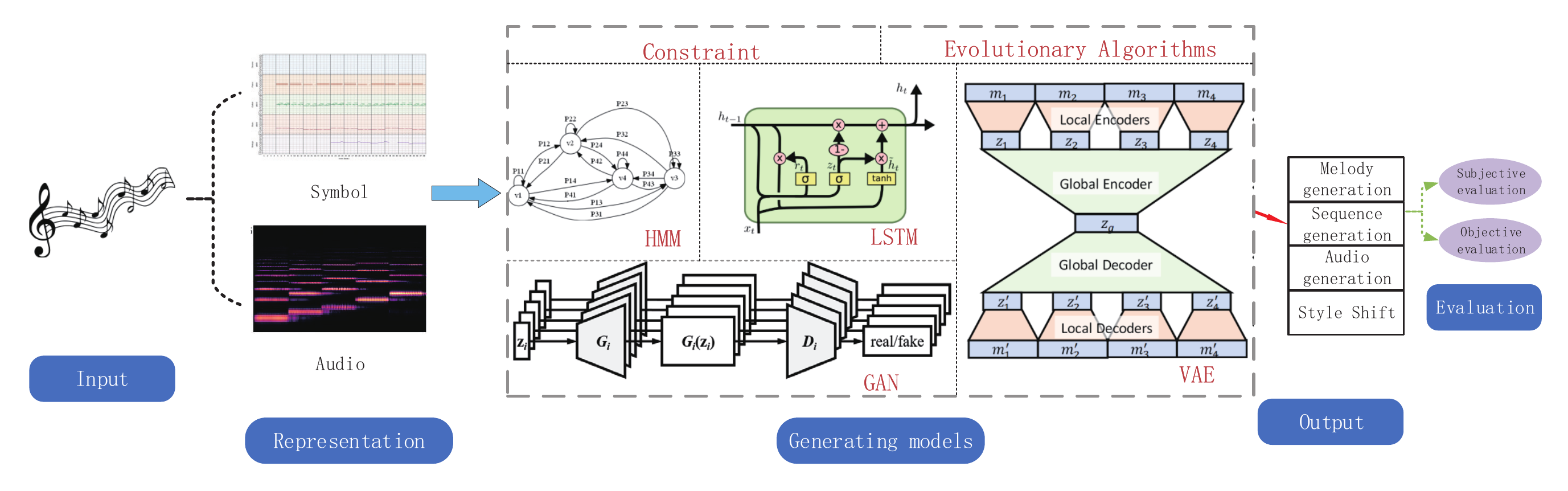}
\caption{Overview of music generation system.}\label{fig_overview}
\end{figure*}

\textcolor{black}{As an area exploring automated creation, computer creativity has driven high-quality research in the direction of intelligent innovation,} including storytelling\cite{avdeeff_artificial_2019} and art\cite{plut_generative_2020}. With the development of artificial intelligence, music has proven to have a very high capacity and potential for automatic creation. Intelligent music generation is dedicated to solving the problem of automatic music composition and is currently one of the most productive sub-fields in the field of computational creativity. 

\textcolor{black}{Symbolic music generation refers to representing music as a sequence of symbols, followed by feature learning and generative modeling\cite{ji2023survey}. Two essential properties of symbolic music composition are structural awareness and interpretive ability.} Structure-aware models can produce naturally coherent music with long-term dependencies, including different levels of repetition and variation\cite{jiang_transformer_2020}; interpretive ability translates complex computational models into controllable interfaces for interactive performance\cite{jin_style-specific_2020}. Finally, this field can serve a variety of purposes, such as composers using the technology to develop existing material, to inspire and explore new ways of making and working with music, and listeners can customise or personalise music.

The artificial intelligence composition system is usually called the Music Generation System (MGS). \textcolor{black}{There are basically 5 steps in music production: 1) composition; 2) arrangement; 3) sound design; 4) mixing and 5) mastering. To our knowledge, contemporary AI music generation encompasses a broad spectrum of music production stages, ranging from sparking creative inspiration and assisting with song structure to fully autonomous composition. These AI systems provide valuable tools and resources for music creators, enabling them to explore, innovate, and interact with technology in various aspects of the music creation process. To conclude, our study mainly focuses on composition and arrangement, covering ideas to a sequence of symbols. The music clips created by generative algorithms encompass a variety of different instruments, contingent upon the complexity of the generation system and the specific objectives of music generation\cite{ji2020comprehensive}. Whether monophonic or polyphonic, these segments include nearly all common instruments. Music generation involving instruments such as bass, drums, guitar, piano, violin, string instruments, wind instruments, electronic synthesizers, among others, has been extensively researched and applied\cite{Musenet, donahue_lakhnes_2019, simon_learning_2018}. Concurrently, the methods and tools for music generation are continuously evolving, enabling the computer-generated production of an increasing array of instruments and music genres. However, it is important to note that the generation of music involving various ethnic instruments still faces limitations and requires ongoing refinement and development.}

\textcolor{black}{Currently, a limited number of review papers on music generation systems are available.} Herremans et al.\cite{herremans_functional_2017} introduced the classification method of music generation systems and pointed out problems in music evaluation. Briot et al.\cite{briot_deep_2020} introduced algorithms based on deep learning based music generation system, and put forward corresponding countermeasures for some limitations in this field. Kaliakatsos et al.\cite{kaliakatsos-papakostas_artificial_2020} analyzed and predicted the research fields that promote the development of intelligent music generation in the field of artificial intelligence; Loughran et al.\cite{loughran_evolutionary_2020} aimed the music generation system based on evolutionary calculation has analyzed the categories and potential problems; Plug et al.\cite{plut_generative_2020} introduced the latest research technology of music generation systems in the game field, and analyzed the prospects and challenges of the field. However, in these previous papers, some do not provide sufficient analysis of algorithms and models\cite{herremans_functional_2017, kaliakatsos-papakostas_artificial_2020,loughran_evolutionary_2020}. \textcolor{black}{In Briot's work\cite{briot_deep_2020}, there exists content chapters with weak thematic relevance, with the basic theory of neural networks serving as an example. These chapters have not substantially contributed to an in-depth exposition of the theme, leading to a dispersion of the work's content.} Others have a more traditional and conservative outlook on the future\cite{herremans_functional_2017, kaliakatsos-papakostas_artificial_2020}, and\cite{plut_generative_2020} is more focused on the application level, giving less information on the general method of music generation.

\textcolor{black}{In light of the limitations of previous review work, this paper comprehensively analyses the most recent articles on the various sub-research directions of music generation, points out their respective strengths and weaknesses, and summarises them in a clear categorisation according to the generation method.} In order to better understand the digital character of the music, we also present the two main representations of music as an information stream and list the commonly used datasets For the evaluation of generative effects, the most representative evaluation methods based on both subjective and objective aspects are selected for discussion. Our work concludes with a comparative analysis of the similarities and differences between Eastern and Western research in the field of AI composition based on different cultural characteristics and a look at the further development of the field of music generation as it matures.

\section{Overview}\label{sec2}

The most basic elements of music include the pitch of the sound, the length of the sound, the strength of the sound and the timbre. These basic elements combine with each other to form the melody, harmony, rhythm and tone of the music. In the field of artificial intelligence composition, melody constitutes the primary purpose of an automatic music generation system. In most melody generation algorithms, the goal is to create a melody similar to the selected style, such as generating western folk music\cite{medeot_structurenet_2018} or free jazz\cite{keerti_attentional_2020}. In addition to melody, harmony is another popular aspect of automatic music generation. When generating a harmony sequence, the quality of its output mainly depends on the similarity with the target style. \textcolor{black}{For example, in choral harmony, this similarity is clearly defined through adherence to sound guiding principles\cite{hadjeres_interactive_2017}, whereas in popular music, chord progressions primarily serve as accompaniments to the melody\cite{zhu_xiaoice_2018}.} In addition, audio generation and style conversion are also popular directions in the field of intelligent composition. The overall framework of the specific music generation system is shown in Figure \ref{fig_overview}. It is worth noting that many historical music generation systems do not include deep neural networks or evolutionary algorithms, but generate outputs based solely on musical rule constraints.
\begin{figure}[h]%
\centering
\includegraphics[width=0.5\textwidth]{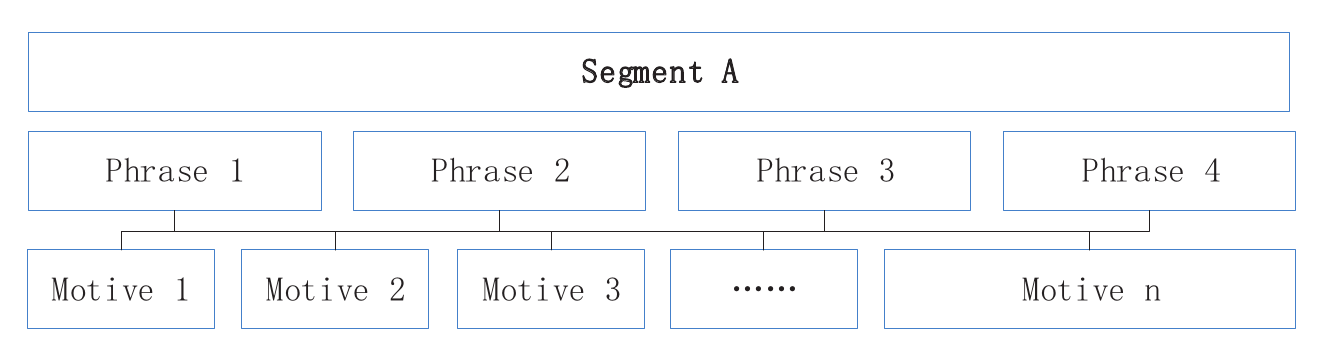}
\caption {The Hierarchy of Music.}\label{fig_hierarchy}
\end{figure}
In the process of generating music, the musical content needs to be encoded in digital form as input to the algorithm, after which the intelligent algorithm is learnt and trained to eventually output different types of musical segments, such as melodies, chords, audio, style transitions, etc... Music has a clear hierarchical structure in time, from motives to phrases and then to segments. As shown in Figure \ref{fig_hierarchy}, where the overall structure of the music together with local repetitions and variants form the theme, rhythm and emotion of the piece; secondly, music usually consists of multiple voices (may be the same or different instruments), each individual voice part has a complete structure, and together they must be in harmony with each other. Combining these limitations, music generation is still a very challenging problem.

\section{Representation}\label{sec3}
\textcolor{black}{Music can be seen as a stream of information that conveys emotions at different levels of abstraction\cite{schafer2013psychological}.} The input to a generative system is usually some representation of music, which has an important influence on the number of input and output nodes (variables) to the system, the correctness of training, and the quality of the generated content\cite{briot_deep_2020}.

\subsection{Specific formats}\label{subsec1}
The representation of music is mainly divided into two categories: audio and symbolic. The main difference between the two is that audio is a continuous signal, while the symbol is a discrete signal.

\subsubsection{Audio representation}\label{subsubsec1}
The main ways in which music can be represented in audio are:

a. Waveform\cite{oord_wavenet_2016}

b. Spectrogram\cite{lu_play_2019}

Among them, the spectrogram\cite{lu_play_2019} can be obtained by the short-time Fourier transform of the original audio signal. The advantage of using audio signal representation is that the original audio waveform retains the original properties of music and can be used to produce expressive music\cite{manzelli_end_2018}. However, this form of representation has certain limitations, because music fragments in the original audio domain are usually represented as continuous waveform $x\in [-1,1]^{T}$, where the number of samples T is the audio duration and the sampling frequency (usually 16 kHz To 48 kHz), short-term music fragments will generate a lot of data. Therefore, modelling the raw audio can lead to a somewhat range-dependent system, making it challenging for intelligent algorithms to learn the high-level semantics of music, such as the Jukebox\cite{dhariwal_jukebox_2020} automated record jukebox that takes approximately nine hours to render one minute of music.

\subsubsection{Symbolic representation}\label{subsubsec2}
The symbolic representation of music fragments is mainly as follows:

a. MIDI events

\begin{figure}[!h]
\centering                                                                   
\subfloat[\textcolor{black}{MIDI events(Extract from Twinkle Twinkle Little Star).}]
{\label{fig_rep.a}
\includegraphics[width=0.5\textwidth]{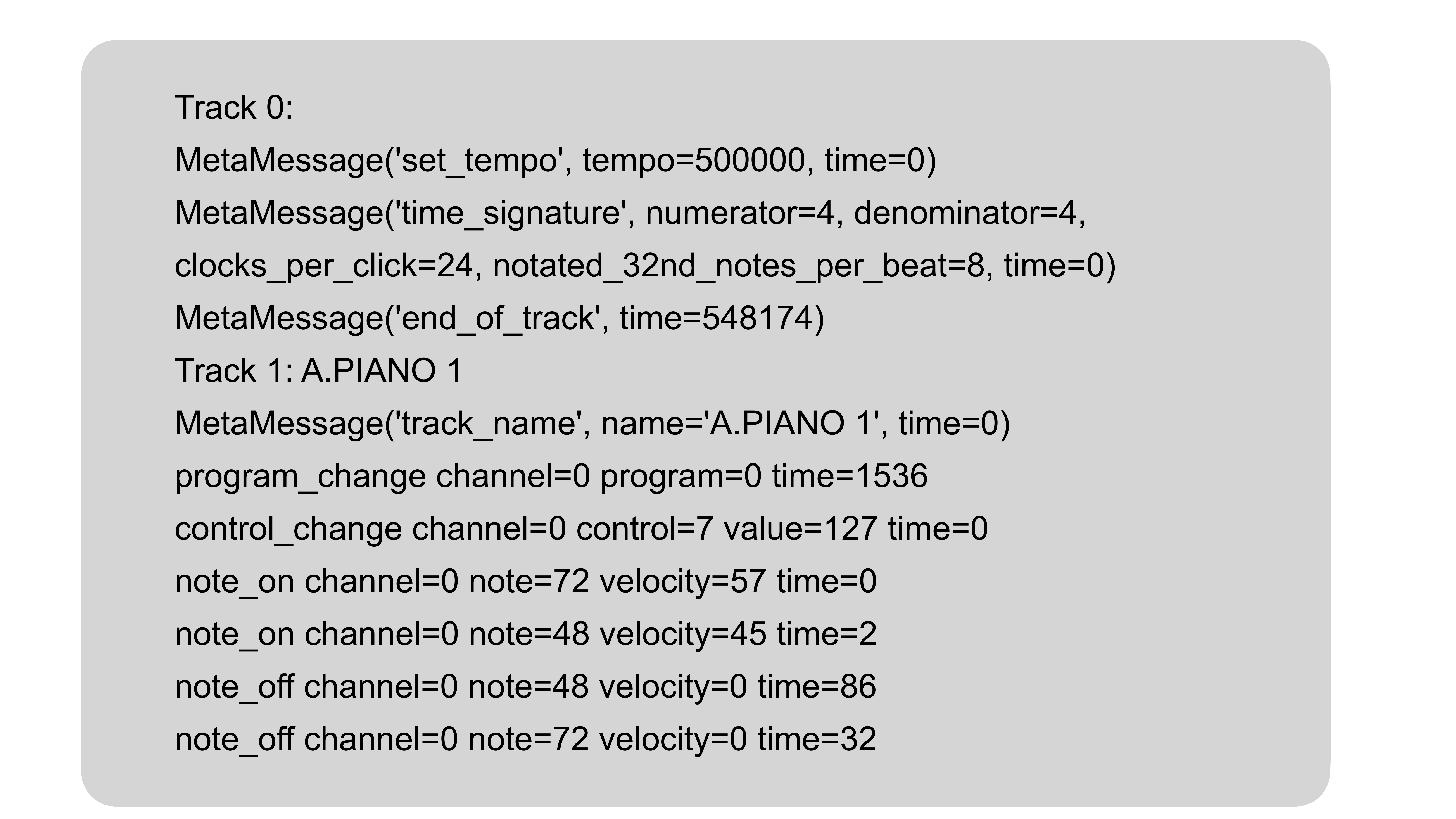}
}
\quad
\subfloat[Piano-roll.]
{\label{fig_rep.b}
\includegraphics[width=0.5\textwidth]{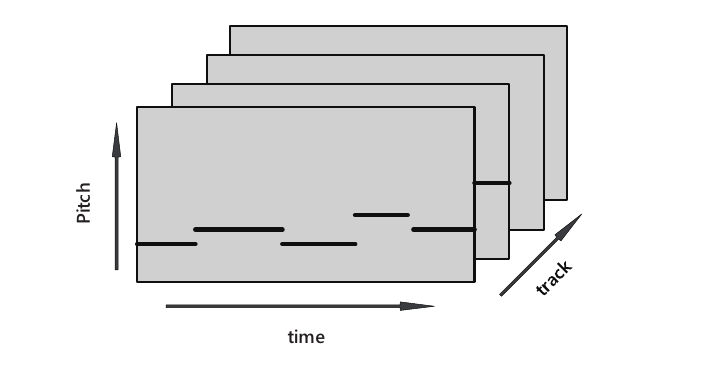}
}
\quad
\subfloat[ABC notation(Extract from Scotland The Blave).]
{\label{fig_rep.c}
\includegraphics[width=0.5\textwidth]{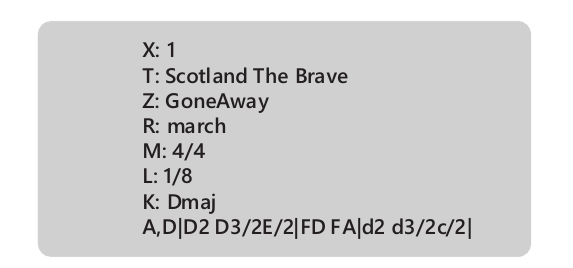}
}
\caption{Representation of Music}\label{fig3}
\end{figure}

MIDI (Musical Instrument Digital Interface) is a digital interface of musical instruments to ensure the interoperability between various electronic musical instruments, software and equipment. Figure \ref{fig_rep.a} shows the type of MIDI event. This method uses two symbols to indicate the appearance of notes: Note on and Note off, indicating respectively the beginning and end of the played notes. In addition, the note number is used to indicate the pitch of the note, which is specified by an integer between 0 and 127. This data structure contains an incremental time value to specify the relative interval time between notes. For example, Music transformer\cite{huang_music_2018}, MusicVAE\cite{roberts_hierarchical_2018}, etc. convert melody, drums, piano, etc. into MIDI events and use them as input to the training network to generate music fragments. However, MIDI events cannot effectively retain the concept of multiple tracks playing multiple notes at once\cite{huang_deep_2016}.

b.Piano-roll

The conversion of a music clip to Piano-roll format begins with a one-hot encoding of the MIDI file before converting it to image form. \textcolor{black}{As shown in Figure \ref{fig_rep.b}, the x-axis represents the continuous time step, the y-axis represents the pitch, and the z-axis represents the MIDI track, where n tracks have n one-hot matrices.}

The Piano-roll format quantizes the notes into a matrix form according to the beats. This representation form is popular in music generation systems because it is easy to observe the structure of the melody and does not require serialization of polyphonic music. \textcolor{black}{The Piano-roll representation is one of the most prevalent methods for representing melodies. For instance, systems like Midi-VAE\cite{brunner_midi-vae_2018}, MuseGAN\cite{dong_musegan_2018}, typically start by converting polyphonic music into the Piano-roll format before proceeding with further processing.} However the limitation of using Piano-roll is that it cannot capture the vocal and other sound details such as timbre, velocity, and expressiveness\cite{briot_deep_2020}, and it is difficult to distinguish long notes from repeated short notes (e.g. whole notes and multiple eighth notes) and has multiple zero integers resulting in low training efficiency.

c.ABC notation

ABC notation$\footnote{http://abcnotation.com/wiki/abc:standard}$ encodes melodies into a textual representation, as shown in Figure \ref{fig_rep.c}, initially designed specifically for folk and traditional music originated in Western Europe, but later widely used for other types of music. A typical example of this encoding method, which has the advantage of being concise and easy to read, is Folk-rnn\cite{sturm_music_2016}, which takes the ABC notation of a particular folk music transcription and uses it as input to an LSTM network to generate tunes in the folk music genre. Similarly, GGA-MG\cite{farzaneh_gga-mg_2020} uses ABC symbols for representation used to generate melodies. \textcolor{black}{However, the application of this representation is limited by the fact that it encodes mainly monophonic melodies.\cite{ji2023survey} Lacking specific notation for polyphony or harmony, it does not excel at encoding complex polyphonic music.}

\subsection{Datasets}\label{subsec2}
\textcolor{black}{For deep neural network-based generative systems, beyond determining the appropriate musical representation, the  selection of a suitable music dataset as the system's input is equally paramount.} Popular public datasets contain music types such as multi-track, choral, piano and folk music, and representations such as audio, MIDI events and ABC symbols. Table \ref{tab1} lists the types and encoding forms of several typical publicly available music datasets for statistical purposes.

\begin{table}[h]
\caption{Examples of music datasets}
\label{tab1}
\begin{tabular}{@{}lll@{}}
\toprule
{\color[HTML]{329A9D} \textbf{Dataset}} & {\color[HTML]{329A9D} \textbf{Type}} & {\color[HTML]{329A9D} \textbf{Format}} \\ \midrule
\textbf{Nsynth}\footnotemark[2]            & Multitrack         & WAV  \\
\textbf{Lakh MIDI Dataset}\footnotemark[3] & Multitrack         & MIDI \\
\textbf{JSB-Chorales}\footnotemark[4]      & Harmonized chorale & MIDI \\
\textbf{Groove MIDI}\footnotemark[5]       & Drum               & MIDI \\
\textbf{Nottingham}\footnotemark[6]        & Folk tunes         & ABC  \\ \bottomrule
\end{tabular}
\end{table}
\footnotetext[2]{https://black.tensorflow.org/datasets/nsynth}
\footnotetext[3]{https://colinraffel.com/projects/lmd/}
\footnotetext[4]{http://www.jsbchorales.net/index.shtml}
\footnotetext[5]{https://black.tensorflow.org/datasets/groove}
\footnotetext[6]{http://abc.sourceforge.net/NMD/}

Nsynth contains 305,979 notes, each with a unique pitch, timbre and envelope. For the 1,006 instruments from the commercial sample library, the four-second, single-single instrument is generated in the range of each pitch (21-108) and five different speeds (25, 50, 75, 100, 127) of a standard MIDI piano.

Lakh MIDI is a collection of 176,581 unique MIDI files, 45,129 of which have been matched and aligned with entries in the One Million Songs Dataset. It aims to facilitate large-scale retrieval of music information, both symbolic (using MIDI files only) and audio content-based (using information extracted from MIDI files as annotations for matching audio files).

JSB-Chorales offers three temporal resolutions: quarter, eighth and sixteenth. These 'quantizations' are created by retaining the tones heard on a specified time grid. Boulanger-Lewandowski (2012) uses a temporal resolution of quarter notes. This dataset does not currently encode fermatas and is unable to distinguish between held and repeated notes.

Groove MIDI Dataset (GMD) consists of 13.6 hours of unified MIDI and (synthetic) audio of human-performed, rhythmically expressive drum sounds. The dataset contains 1150 MIDI files and over 22,000 drum sounds.

Nottingham Music Dataset maintained by Eric Foxley contains over 1000 folk tunes stored in a special text format. Using NMD2ABC, a program written by Jay Glanville and some perl scripts, much of this database has been converted to ABC notation.

\textcolor{black}{Based on a comprehensive survey of existing work, it is evident that music generation datasets, represented by the aforementioned datasets, predominantly focus on the collection and processing of Western music genres, with relatively limited attention given to Eastern music genres. While some researchers have gathered and processed music from Eastern genres, it is regrettable that only a few of these datasets are accessible.} MG-VAE\cite{luo_mg-vae_2019} digitized the folk songs from various regions of China as documented in "Chinese Folk Music Collection," converting them into 2000 distinct regional-style compositions presented in MIDI format. Moreover, the dataset curated by XiaoIce Band\cite{zhu_xiaoice_2018} focuses on Chinese pop music, suitable for generating multi-track pop music compositions. Jiang et al.\cite{jiang_stylistic_2019} compiled a dataset comprising over a thousand Chinese karaoke songs, and the CH818\cite{hu_mood_2017} dataset contains 818 Chinese Xpop (Chinese pop music) songs, from each of which 30-second segments evoking the strongest emotions were extracted and appended with emotional labels. \textcolor{black}{Ajay et al.\cite{srinivasamurthy2021saraga} introduce two large open data collections of Indian Art Music which currently form the largest annotated data collections available for computational analysis of Indian Art Music.}

\textcolor{black}{However, it is worth noting that despite the existence of these Eastern music datasets, their accessibility is restricted, possibly due to copyright-related concerns and other issues. To promote the advancement of the relevant research field, further concerted efforts are required to actively construct more comprehensive and diversified datasets of Eastern music, better catering to the research needs. And this necessitates extensive international collaboration, cultural compliance, data transparency, and technological innovation. By enhancing the diversity of datasets, promoting open sharing, utilizing data augmentation techniques, and harnessing community participation, it is possible to provide richer resources for research in Eastern music generation, thereby advancing understanding and facilitating the creative and cultural preservation of Eastern music.}

\section{Music Generation Systems}\label{sec4}
The quest for computers to create music has been pursued by researchers since the dawn of computing. \textcolor{black}{In 1957, Lejaren Hiller produced the IlliacSuite \cite{hiller1957musical}, the melody of its fourth movement was the first musical composition in human history composed entirely by a computer using Markov chain.} David Cope(1991)\cite{david_cope} introduced Experiments in Musical Intelligence (EMI), which used the most basic process of pattern recognition and recombinancy to create music that successfully imitated the styles of hundreds of composers.

\textcolor{black}{In the contemporary era, the trajectory of music generation systems has traversed a developmental continuum, transitioning from early rule-based methodologies\cite{supper_few_2001} to the presently ubiquitous deep learning-based paradigms\cite{briot_deep_2020}. This evolution is characterized by a gradual shift in the generated musical output, progressing from rudimentary and brief monophonic musical fragments to intricate polyphonic compositions that incorporate harmonious chords and protracted durations.} This paper surveys the recent state-of-the-art in music generation and classifies them based on algorithm, genre, dataset, and representation, etc. The detailed classification results are presented in Table \ref{tab2}, which can be simply categorized into the following categories according to the purpose of generation:

\begin{itemize}
\item Melody generation
\item Arrangement generation
\item Audio generation 
\item Style transfer
\end{itemize}

% \newpage
\textcolor{black}{The melody generator can produce melodies for instruments such as drums and bass, while the composition generator is capable of generating harmonies, lead melodies, and counterpoint, among others.} Audio generation involves modeling music directly at the audio level, generating sound segments with different instruments and styles. This method presents challenges, including computational complexity and the difficulty of capturing semantic structure from raw audio. Style transfer encompasses timbre conversion, arrangement style transformation, etc., and attempts to apply a specific style (such as a composer or musical genre) or a particular instrument (e.g., drums, bass) to the original musical composition using automatic arrangement techniques. Detailed classification information for each music generation system can be found in Tab \ref{tab2}. Although many music generation algorithms feature combinations and nesting, such as the use of Variational Auto-Encoder (VAE) combined with Recursive Neural Networks (RNN)\cite{roberts_hierarchical_2018}, music generation algorithms can be generally categorised as follows:

\begin{itemize}
\item Rule Based Systems
\item Markov Model
\item Deep Learning: LSTM, VAE, GAN et al.
\item Evolutionary Computation, EC
\end{itemize}

\onecolumn
\begin{landscape}
\begin{center}
\begin{longtable}{@{}lllll@{}}
\caption{List of Music Generation Systems}
\label{tab2}\\
\multicolumn{1}{c}{{\color[HTML]{2E74B5} \textbf{}}} & Name                              & Year                   & Algorithm                                                       & Application                           \\* \midrule
\endfirsthead
\endhead
\bottomrule
\endfoot
\endlastfoot
Rule                                                 & T. Tanaka et al.              & 2015                   & Constraint based System                                         & melody generation                     \\
                                                     & MorpheuS                      & 2017                   & Constraints(tonal and pattern)                                  & Arrangement generation                \\*  \midrule
                                                     & Chih-Fang Huang et al.        & 2016                   & Markov chain                                                    & melody generation                     \\
                                                     & AS   Ramanto et al.           & 2017                   & Markov chain                                                    & melody generation                     \\
\multirow{-3}{*}{Markov Model}                       & D   Williams et al.           & 2019                   & Hidden Markov Model(HMM)                                        & melody generation                     \\* \midrule
                                                     & Wavenet[13]                       & 2016                   & Dilated Casual Convolutions                                     & Audio generation                      \\
                                                     & Engel   et al.                & 2017                   & Wavenet/ AE                                                     & timbre style transfer                 \\
                                                     & Manzelli   et al.             & 2018                   & WaveNet/biaxial LSTM                                            & Audio generation                      \\
                                                     & TimbreTron                    & 2019                   & WaveNet/CycleGAN                                                & composition style transfer            \\
                                                     & Coconet                       & 2019                   & Counterpoint by Convolution                                     & Reconstruct partial scores            \\
\multirow{-6}{*}{CNN}                                & PerformanceNet                & 2019                   & ContourNet/TextureNet                                           & Audio generation                      \\* \midrule
                                                     & MelodyRNN                     & 2016                   & lookback RNN/attention RNN                                      & melody generation                     \\
                                                     & Hadjeres   et al.             & 2017                   & Constraint-RNN/Token-RNN(LSTM)                                  & melody generation                     \\
                                                     & RL   Tuner                    & 2017                   & LSTM/RL                                                         & melody generation                     \\
                                                     & StructureNet                   & 2018                   & LSTM                                                            & melody generation                     \\
                                                     & Makris   et al.               & 2019                   & LSTM/FF                                                         & melody generation                     \\
                                                     & Keerti   et al.               & 2020                   & bi-LSTM/attention                                               & melody generation                     \\
                                                     & Folk-rnn                      & 2016                   & LSTM                                                            & Arrangement generation                \\
                                                     & Song   From PI                & 2016                   & LSTM                                                            & Arrangement generation                \\
                                                     & JamBot                        & 2017                   & LSTM                                                            & Arrangement generation                \\
                                                     & DeepBach                      & 2017                   & bi-LSTM/pseudo-Gibbs sampling                                   & Arrangement generation                \\
                                                     & XiaoIce   Band                & 2018                   & GRU/MLP/multi-task learning                                     & Arrangement generation                \\
                                                     & Amadeus                       & 2019                   & LSTM/RL                                                         & Arrangement generation                \\
                                                     & Performance   RNN            & 2018                   & LSTM                                                            & Audio generation                      \\
\multirow{-14}{*}{RNN}                               & Jin et   al.                   & 2020                   & LSTM/AC RL                                                      & Composition style transfer            \\* \midrule
                                                     & C-RNN-GAN                     & 2016                   & GAN/LSTM                                                        & melody generation                     \\
                                                     & MidiNet                      & 2017                   & GAN/CNN                                                         & melody generation                     \\
                                                     & JazzGAN                      & 2018                   & GAN/RA                                                          & melody generation                     \\
                                                     & SSMGAN                        & 2019                   & GAN/ SSM                                                        & melody generation                     \\
                                                     & Yu et   al.                   & 2019                   & GAN/conditional LSTM                                            & melody generation                     \\
                                                     & MuseGAN                       & 2018                   & GAN                                                             & Arrangement generation                \\
\multirow{-7}{*}{GAN}                                & BinaryMuseGAN                & 2018                   & GAN/ BN                                                         & Arrangement generation                \\* \midrule
                                                     & LeadSheetGAN                  & 2018                   & GAN/CNN                                                         & Arrangement generation                \\
                                                     & CycleBEGAN                   & 2018                   & CycleBEGAN-CNN/skip/recurrent                                   & Timbre style transfer                 \\
                                                     & CycleGAN                      & 2018                   & Cycle-GAN                                                       & Composition style transfer            \\
                                                     & Play as   You Like            & 2019                   & RaGAN/MUNIT                                                     & composition style transfer            \\
                                                     & WaveGAN                       & 2019                   & DCGAN                                                           & audio generation                      \\
\multirow{-6}{*}{GAN}                                & GANSynth                      & 2019                   & PGGAN                                                           & audio generation                      \\* \midrule
                                                     & MusicVAE                      & 2019                   & VAE/RNN                                                         & melody generation                     \\
                                                     & GrooVAE                       & 2019                   & VAE/LSTM                                                        & melody generation                     \\
                                                     & Jiang   et al.                & 2019                   & attention-VAE                                                   & melody generation                     \\
                                                     & MuseAE                        & 2020                   & AAE/bi-LSTM                                                     & melody generation                     \\
                                                     & MIDI-Sandwich2                & 2019                   & BVAE/MFG-VAE                                                    & Arrangement generation                \\
                                                     & Rivero   et al.               & 2020                   & VAE                                                             & Arrangement generation                \\
                                                     & Jukebox                       & 2020                   & VQ-VAE                                                          & Arrangement generation                \\
                                                     & MIDI-VAE                      & 2018                   & parallel VAE/GRU                                                & Composition style transfer            \\
                                                     & MahlerNet                     & 2019                   & conditional VAE/bi-RNN                                          & Composition style transfer            \\
\multirow{-10}{*}{VAE}                               & MG-VAE                        & 2019                   & VAE/bi-GRU                                                      & Composition style transfer            \\* \midrule
                                                     & Music Transformer             & 2019                   & transformer                                                     & Arrangement generation(piano)         \\
                                                     & LakhNES                       & 2019                   & transformer XL                                                  & Arrangement generation                \\
                                                     & MuseNet                       & 2019                   & Transformer                                                     & Arrangement generation                \\
                                                     & Guan et   al.                 & 2019                   & self-attention/GAN                                              & Arrangement generation                \\
                                                     & Zhang   et al.                & 2020                   & transformer/GAN                                                 & Arrangement generation                \\
                                                     & Transformer   VAE              & 2020                   & transformer/VAE                                                 & melody generation                     \\
\multirow{-7}{*}{Transformer}                        & Choi et   al.                 & 2020                   & Transformer/VAE                                                 & melody generation                     \\* \midrule
                                                     & Muńoz et al.                  & 2016                   & MA/fitness:Harmonic rules                                       & melody generation                     \\
                                                     & Jeong et   al.                & 2017                   & EA/fitting function:   Multi-objective (stability, tension)     & melody generation                     \\
                                                     & GGA-MG                       & 2020                   & GA/fitting function: LSTM   network                             & melody generation                     \\
                                                     & S   Hickinbotham et al.       & 2016                   & GA/fitness function: Interactive                                & Live-coding                           \\
                                                     & Kaliakatsos   et al.          & 2016                   & EA/PSO                                                          & Arrangement   generation(interactive) \\
                                                     & Olseng   O et al.             & 2018                   & EA/ fitness   function:multi-objective                          & Arrangement generation                \\
                                                     &                                   &                        & EA/ fitness   function:multi-objective                          & Arrangement generation                \\
\multirow{-10}{*}{EA}                                & \multirow{-2}{*}{EvoComposer} & \multirow{-2}{*}{2019} & Harmony   rules, corpus statistical analysis                    & Arrangement generation                \\* \bottomrule
\end{longtable}
\end{center}
\end{landscape}

\twocolumn

\subsection{Rule Based Music Generation}\label{subsec3}
\begin{figure}[h]%
\centering
\includegraphics[width=0.45\textwidth]{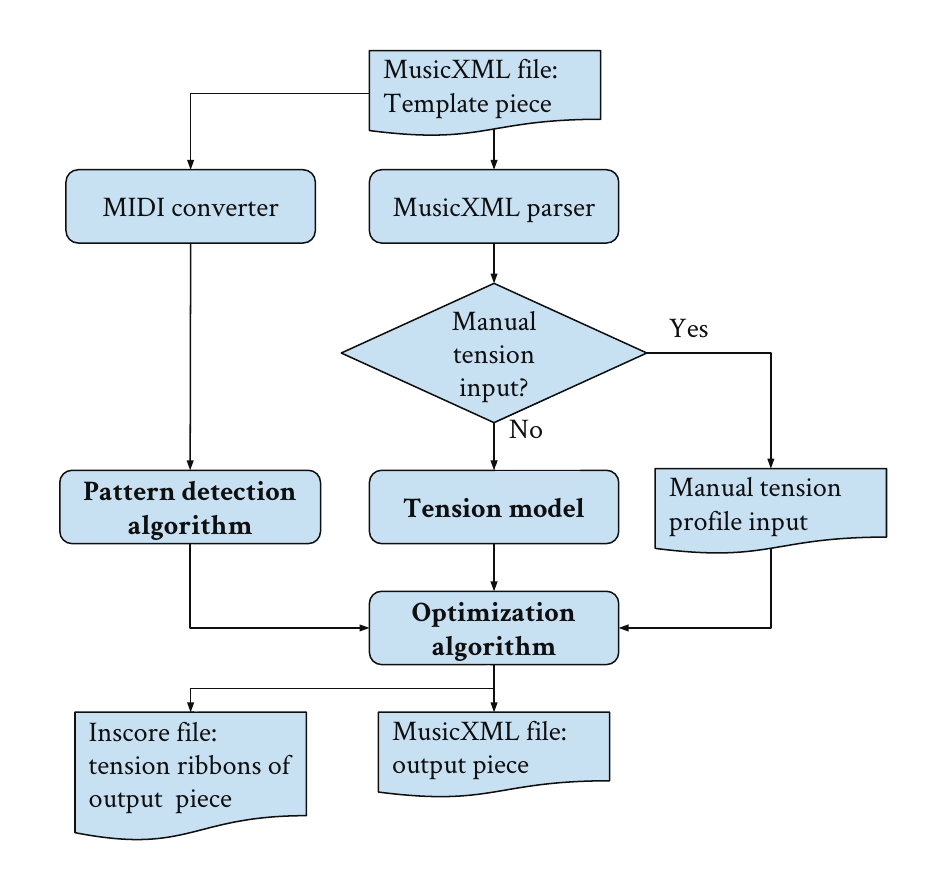}
\caption{\textcolor{black}{Overview of MorhpeuS' architecture\cite{herremans_morpheus_2019}.}}\label{fig_constrain_based}
\end{figure}
The rule based music generation system is a non-adaptive model, which generates music based on the theoretical knowledge of music. Including rule based on grammar\cite{supper_few_2001} and its variant L-system, rule-based constraints: such as similarity\cite{leach_nature_1995}, rhythm rules \cite{herremans_morpheus_2019}, etc. \textcolor{black}{These constraints help guide the process and direction of intelligent music composition.}
The performance of a rule based generation system largely depends on the creativity of the developer, the musical concept of the inventor, and how this abstraction is expressed in terms of the corresponding variables\cite{kaliakatsos-papakostas_artificial_2020}. T.Tanaka et al.\cite{supper_few_2001} proposed a formula for generating music rhythm based on grammatical rules, focusing on the structure and redundancy of music, and controlling some global structure of the note rhythm sequence, but the project is still a theoretical expression, has not yet been practiced. Miguel et.al.\cite{delgado2009inmamusys} introduced the user's emotional input based on the use of rules derived from music theory knowledge and proposed a bottom-up approximation approach to design a two-level multi-agent structure for music generation in a modular way. Although the problem of musical expressiveness is well solved, the system is still at a rather early stage of development, handling user input very rigidly and performing poorly when dealing with fuzzy concepts.  Besides, as figure \ref{fig_constrain_based} shows, MorpheuS\cite{herremans_morpheus_2019} uses Variable Neighborhood Search (VNS) to assign pitches by detecting repeated rhythms in music to generate polyphonic music with specific tension and repeating rhythms.

Rule based generative systems can effectively constrain the exploration space. However,  music is organized on multiple abstract levels,  even in highly structured and rule-bound compositions like Bach chorales\cite{huang_counterpoint_2019}, complex rules may not be sufficient to capture deeper-level structures. Rule  based music generation systems to some extent limit the diversity of their outputs.  As the system involves more analysis and rules, it tends to introduce higher degrees of ambiguity.
\subsection{Markov Model}\label{subsec4}

Markov model\cite{huang_analyzing_2016, ramanto_markov_2017, williams_ai_2019} is a probability-based generation model for processing time series data, i.e. data where there is a time series relationship between samples. The probability of occurrence of specific elements in the data set and the probability of well-defined features can be captured, such as the occurrence of notes or chords, the transition of notes or chords, and the conditional probability of generating chords on a given note.

\begin{figure}[h]%
\centering
\includegraphics[width=0.45\textwidth]{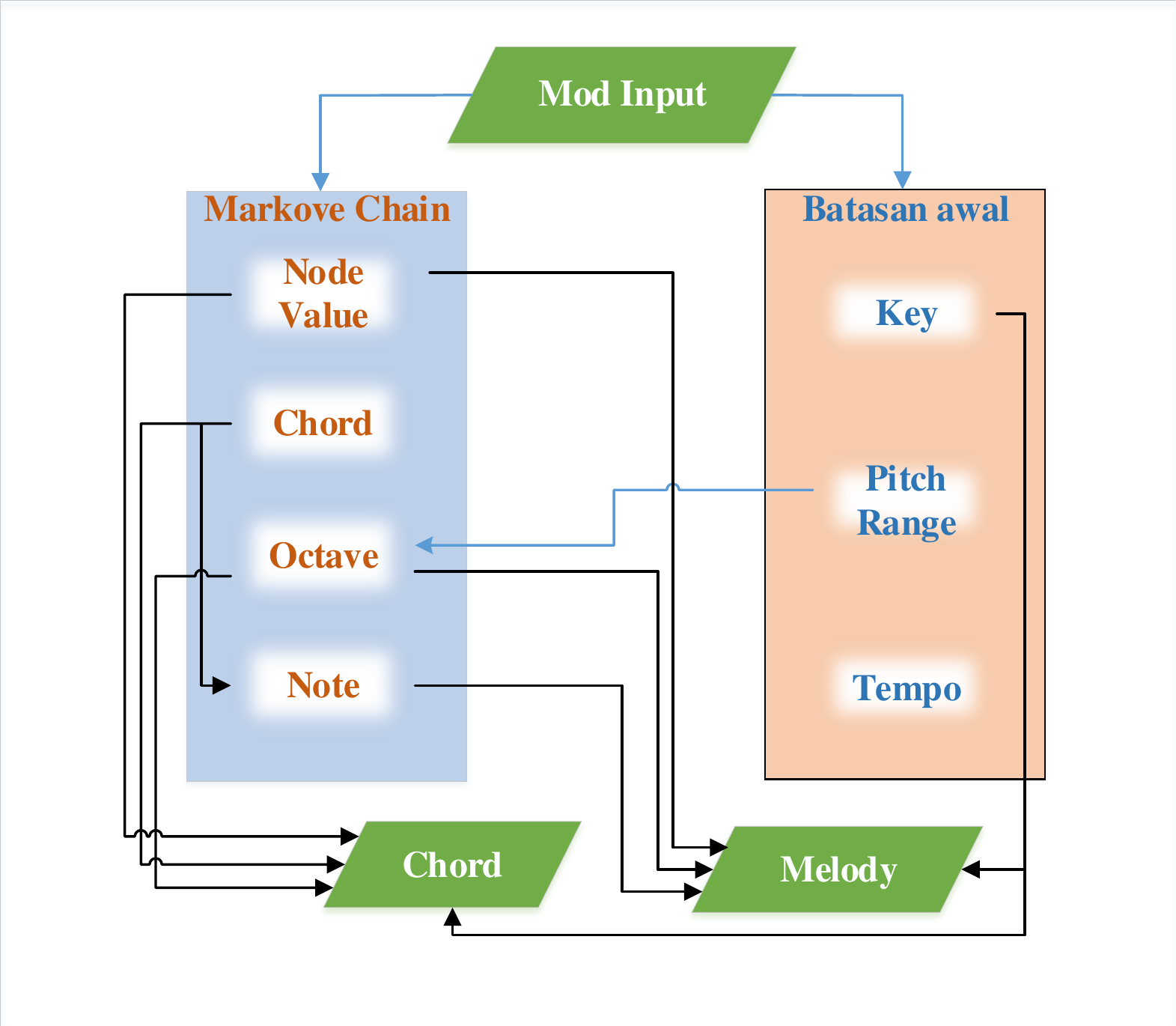}
\caption{Programmable Markov Music Generation System\cite{ramanto_markov_2017}.}\label{fig_markov}
\end{figure}

\textcolor{black}{Due to the temporal nature of music segments, Markov models are highly suitable for generating music elements such as melodies (note sequences)\cite{creavitity}.}Chi-Fang Huang et al.\cite{huang_analyzing_2016} established a Markov chain switching table of the Chinese pentatonic scale by analyzing the modes of Chinese folk music. They generate the next note of the current input melody according to the Markov chain switching table and the rhythm complexity algorithm, and re-present a specific style of music. A S Ramanto et al.\cite{ramanto_markov_2017} used four Markov chains to model the components of a musical fragment including tempo, pitch, note value and chord type, and assigned parameter values for different moods to each part, thus ensuring that the music generation system could adapt to different moods of the input. The model of the procedural music generation algorithm used in the literature is shown in Figure \ref{fig_markov}. D Williams et al.\cite{williams_ai_2019} also generated melodies based on emotions, with the difference that the approach used a Hidden Markov Model (HMM) for modelling, showing that emotionally relevant music can be composed using a Hidden Markov Model.

Compared with the deep learning frame, the advantage of the Markov model is that it is easier to control, allowing constraints to be attached to the internal structure of the algorithm\cite{briot_deep_2020}. \textcolor{black}{However, from a creative perspective, Markov models may yield more novel combinations for shorter sequences, but for longer, structured sequences, they might non-innovatively reuse many elements from the corpus, leading to an excessive repetition of segments.}

\subsection{Deep Learning Methods for Music Generation}\label{subsec5}
Content generation is an extended area of deep learning. With the achievements of \textcolor{black}{Google's black\footnotemark[7] and CTRL (Creator Technology Research Lab)\footnotemark[8] in music generation, deep learning as a method for music generation is garnering increasing attention.} Unlike grammar-based or rule-based music generation systems, deep learning based music generation systems can learn the distribution and relevance of samples from an arbitrary corpus of music and generate music that represents the musical style of the corpus through prediction (e.g. predicting the pitch of the next note of a melody) or classification (e.g. identifying the chords corresponding to the melody).

\footnotetext[7]{https://black.tensorflow.org/}
\footnotetext[8]{https://www.francoispachet.fr/}

\subsubsection{Convolutional Neural Networks}\label{subsubsec3}
\begin{figure}[h]%
\centering
\includegraphics[width=0.5\textwidth]{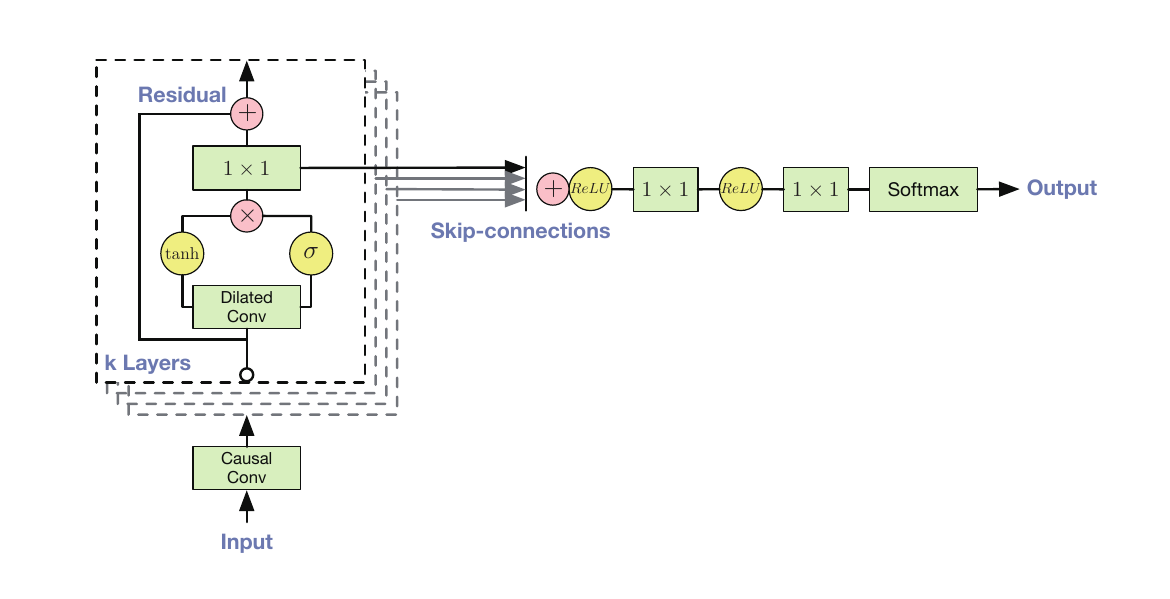}
\caption{\textcolor{black}{WaveNet: Overview of the residual block and the entire architecture\cite{oord_wavenet_2016}}.}\label{fig_convolutional}
\end{figure}
\textcolor{black}{Convolutional neural networks (CNNs) are primarily used for processing image data that does not involve time-series characteristics, making it relatively challenging to handle music data with time-series features. As a result, there are currently fewer music generation models based on CNNs. However, it is still feasible to generate music through improvements in model architecture or sophisticated data preprocessing techniques.} \par
The audio generation model WaveNet\cite{oord_wavenet_2016}, introduced by Google Deepmind, including its variants TimbreTron\cite{manzelli_conditioning_2018} (timbre conversion), Manzelli et al.\cite{engel_neural_2017} (audio generation) provide effective improvements to convolutional neural networks to enable them to generate music clips.  As figure \ref{fig_convolutional} shows, Wavenet generates new music clips with high fidelity by introducing extended causal convolution (Dilated Casual Convolutions), which predicts the probability distribution of the current occurrence of different audio data based on the already generated audio sequences. By introducing extended convolution, the deep model has a very large perceptual field, thus dealing with the long-span time-dependent problem of audio generation. However, there is no significant repetitive structure in the generated music fragments and the generated audio is less musical. Manzell et al. improved this method by using an LSTM (Long Short-Term Memory) network combined with the \textcolor{black}{Wavenet architecture to generate cello-style audio clips based on a given melody from the MusicNet dataset\cite{thickstun2016learning}, such as the Happy Birthday and a C major scale.} Where the LSTM network is used to learn the melodic structure of different musical styles, the output of this model is used as input to the Wavenet audio generator to provide the corresponding melodic information for the generation of the audio clip.

The Wavenet model can also be used for other types of music generation tasks, with Engel et al.\cite{engel_neural_2017} constructing a Wavenet-like encoder to infer hidden temporal distribution information and feeding it into a Wavenet decoder, which effectively reconstructs the original audio and enables conversion between the timbres of different instruments. Similarly, TimbreTron\cite{huang_timbretron_2019} is a musical timbre conversion system that applies 'image' domain style transfer to the time-frequency representation of audio signals. The timbre transfer is performed in the log-CQT domain using CycleGAN (Cycle Generative Adversarial Networks) by first calculating the Constant Q Transform of the audio signal and using its logarithmic amplitude as an image, and finally using the Wavenet synthesiser to generate high-quality waveforms, enabling the transformation of instrument timbres such as cello, violin and electric guitar. In addition, Coconet\cite{huang_counterpoint_2019} used a convolutional alignment network to automate the filling of incomplete scores by modelling the tenor, baritone, bass and soprano in a Bach chorus as an eight-channel feature map, using the feature map of randomly erased notes as input to the convolutional neural network.

\subsubsection{Recurrent Neural Networks}\label{subsubsec4}
\begin{figure}[h]%
\centering
\includegraphics[width=0.45\textwidth]{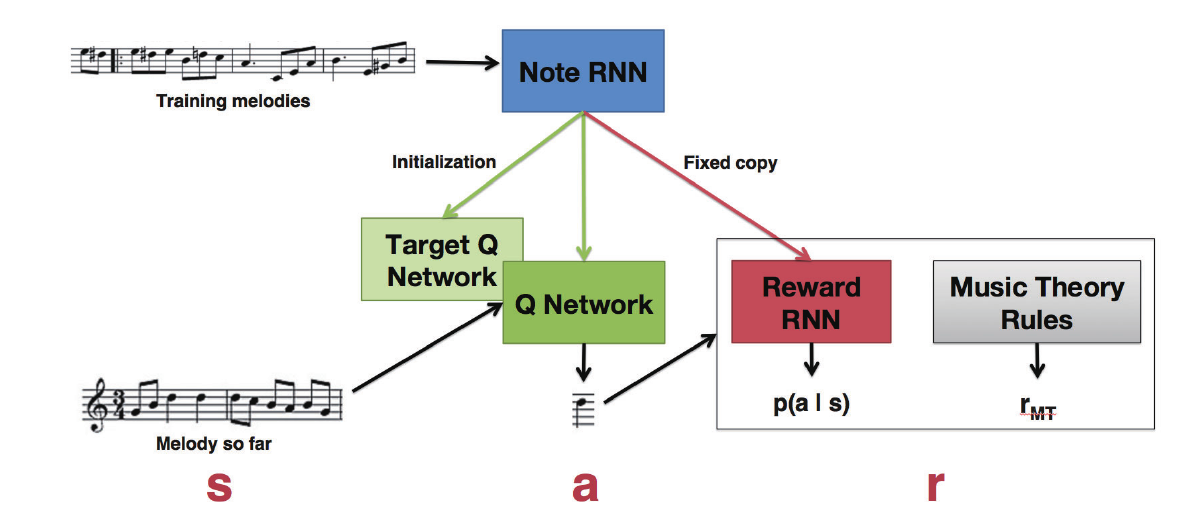}
\caption{A Note RNN is trained on MIDI files and supplies the initial weights for the Q-network
and Target-Q-network, and final weights for the Reward RNN\cite{waite_generating_2016}.}\label{fig_rnn}
\end{figure}
\textcolor{black}{Recurrent Neural Networks (RNN) are a class of neural networks for processing time series, which are very suitable for processing music clips.} However, RNN cannot solve the long time dependency problem due to the gradient problem that occurs when processing long time sequence data, which is prone to gradient explosion (disappearance). LSTM, as a variant of RNN, effectively alleviates the long time dependency problem of RNN by introducing cell states and using three types of gates, namely input gates, forgetting gates and output gates, to hold and control information. LSTM networks are mainly used in music generation systems to generate music clips instead of RNN networks.

The combination of LSTM and other algorithms can effectively generate melodies, polyphonic music, etc. The MelodyRNN (figure \ref{fig_rnn}) proposed by the Google Brain team\cite{waite_generating_2016} built a lookback RNN and an attention RNN in order to improve the ability of RNNs to learn long sequence structures, where the attention RNN introduced an attention mechanism (attention) to model higher-level structures in music to generate melodies. Keerti et al.\cite{keerti_attentional_2020} also introduced an attention mechanism and used dropout to reduce overfitting to generate jazz music. Rl Tuner\cite{jaques_tuning_2017} used an RNN network to provide partial reward values for a Reinforcement Learning (RL) model, with the reward values consisting of two components: (1) the similarity of the next note generated to the note predicted by the reward RNN; (2) the degree of satisfaction with the user-defined constraints. However, this method is based on overly simple melodic composition rules and can only generate simple monophonic melodies. Other music generation systems for melody generation include Hadjeres et al.\cite{hadjeres_interactive_2017}, which proposes an Anticipation-RNN neural network structure for generating melodies for interactive Bach-style choruses; StructureNet\cite{medeot_structurenet_2018} generates simple monophonic accompanying music based on LSTM networks; and Makris et al.\cite{makris_conditional_2019} constructed a separate LSTM network for each drum type, with a feed-forward (FF) network serving as the conditional layer. Its input consists of a one-hot matrix containing 26 features, including rhythm and time signature, and it can generate sequences for drums without prior training on time signatures.

Regarding arrangement generation, Folk-rnn first used LSTM networks to model musical sequences transcribed into ABC notation for generating folk music. DeepBach\cite{hadjeres_deepbach_2017} used a bi-directional LSTM (Bi-LSTM) to generate Bach-style choral music considering two directions in time: one aggregating past information and the other aggregating future information, demonstrating that the sequences generated by the bi-directional LSTM were more musical. XiaoIce Band\cite{zhu_xiaoice_2018} proposes an end-to-end Chinese multi-track pop music generation framework that uses a GRU network to handle the low-dimensional representation of chords and obtains hidden states through encoders and decoders, ultimately generating music for multi-instrument tracks. Amadeus\cite{kumar_polyphonic_2019} uses an explicit duration encoding approach: multiple audio streams represent note durations and uses RL's reward mechanism to improve the structure of the generated music. Jambot\cite{brunner_jambot_2017} employs a chord LSTM network to predict chord sequences, and a polyphonic LSTM network to generate polyphonic music based on the predicted chord sequences. Song From PI\cite{chu_song_2016} uses a layered RNN network model, where the two bottom LSTM networks generate the melody and the two top LSTM networks generate the drums and chords.

In addition, audio generation models such as Performance RNN\cite{oore_this_2020} and PerformanceNet\cite{wang_performancenet_2019} are considered to be the possible future development directions of music generation systems\cite{Dong HW}. Music performance is a necessary condition for giving meaning and feeling to music. Direct synthesis of audio with a library of sound samples often leads to mechanical and dull results. Performance RNN converts a MIDI file of a live piano piece into a musical representation of multiple one-hot vectors with 413 dimensions, and the method specifies a time "step" to a fixed size (10ms) rather than a note time value, thus capturing more expressiveness in the note timing. PerformanceNet is the first attempt at score-to-audio conversion using a fully convolutional neural network with a symbolic representation of the music (Piano-roll) as input and an audio representation (sound spectrum map) as output. The model consists of two sub-networks: a convolutional encoder/decoder structure is used to learn the correspondence between Piano-roll and the acoustic spectrogram; a multi-segment residual network is further used to optimise the results by adding spectral textures for overtones and timbres, ultimately generating music clips for violin, cello and flute. 

\subsubsection{Generative Adversarial Nets}\label{subsubsec5}
\begin{figure}[h]%
\centering
\includegraphics[width=0.45\textwidth]{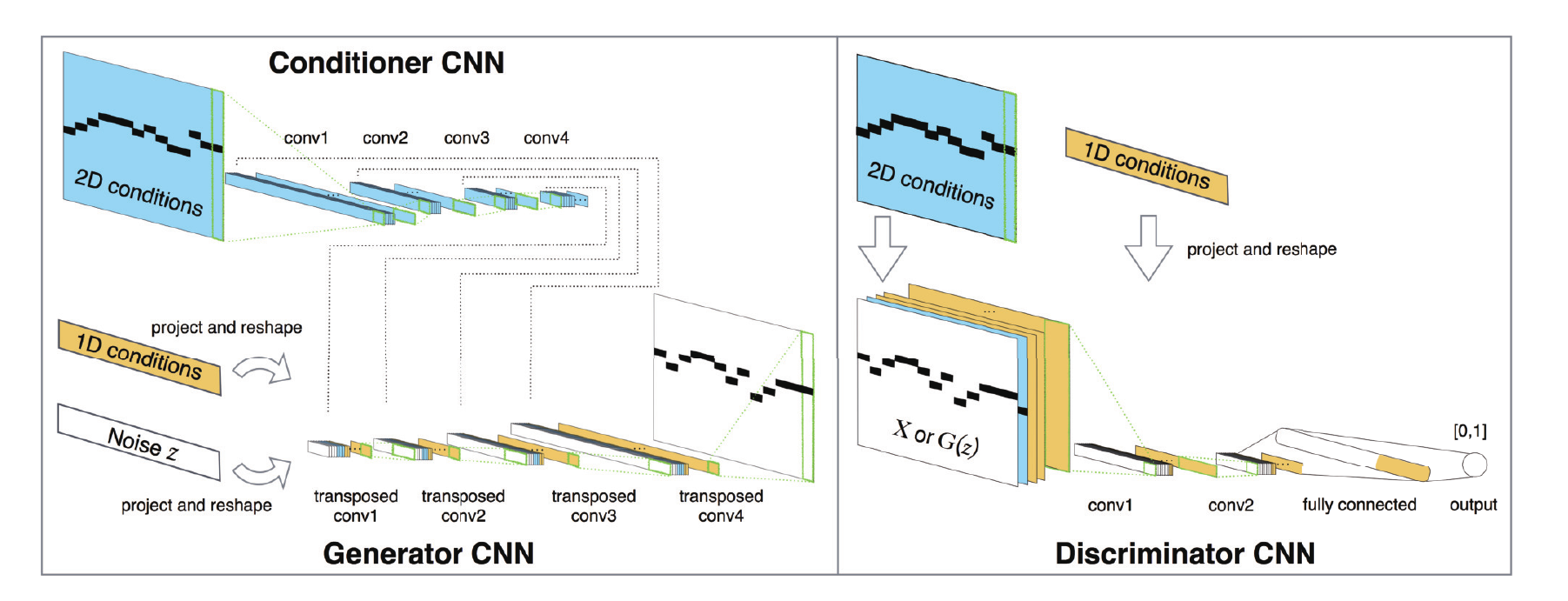}
\caption{\textcolor{black}{System diagram of the MidiNet model for symbolic-domain music generation\cite{yang_midinet_2017}.}}\label{fig_gan}
\end{figure}
Generative Adversarial Nets (GAN)\cite{goodfellow_generative_2014} is a state-of-the-art method for generating high-quality images. The idea behind the method is to train two networks simultaneously: a generator and a discriminator, with the two lattices playing against each other, with the discriminator eventually being unable to distinguish between the truth and falsity of the real data and the generated data as the optimal result. Researchers have been working on applying this to more sequential data, such as audio and music. For music generation, the problems are that:

\begin{enumerate}
\item [(1)]music is of a certain time series
\item [(2)]music is multi-track/multi-instrumental
\item [(3)]monophonic music generation cannot be used directly for polyphonic music
\end{enumerate}

Numerous researchers have improved GAN networks to generate melodies, arrangements, style transformations, etc.

For melody generation, C-RNN-GAN\cite{mogren_c-rnn-gan_2016} uses LSTM networks as a generator and discriminator model, but the method lacks a conditional generation mechanism to generate music based on a given antecedent. MidiNet\cite{yang_midinet_2017} improves the method by inserting the melody and chord information generated in the previous stage as a conditional mechanism in the middle convolution layer of the generator to restrict the generation of note types, thus improving the innovation of generating single jazz pieces, and the diagram of this model is shown in figure\ref{fig_gan}. JazzGAN\cite{trieu_jazzgan_2018} developed a GAN-based model for single jazz generation, using LSTM networks to improvise monophonic jazz music over chord progressions, and proposed several metrics for evaluating the musical features associated with jazz music. SSMGAN\cite{jhamtani_modeling_2019} used the GAN model to generate a self-similar matrix (SSM) to represent musical self-repetition, and subsequently fed the SSM matrix into an LSTM network to generate melodies. Yu et al.\cite{yu_conditional_2021} first proposed a conditional LSTM-GAN for melody generation from lyrics, containing an LSTM generator and LSTM discriminator, both with lyrics as conditional inputs.

For arrangement generation, MuseGAN proposed a GAN model with three types of generators to construct correlations between multiple tracks, containing independent generation within tracks, global generation between tracks, and composite generation, but the method produced many over-fragmented notes. 
BinaryMuseGAN\cite{dong_convolutional_2018} improved upon the aforementioned method by introducing binary neurons (BN) as inputs to the generator. This representation makes it easier for the discriminator to learn decision boundaries. It also introduces chromatic features to detect chord characteristics, reducing the excessive fragmentation of notes. LeadSheetGAN\cite{liu_lead_2018} uses a functional score (Lead Sheet) as a conditional input to generate a Piano- roll, and this approach produces results that contain more information at the musical level due to the clearer melodic direction of the functional score compared to MIDI data.

In terms of style transfer and audio generation, CycleGAN\cite{brunner_symbolic_2018} introduces a style loss function for style conversion and a content loss function to maintain content consistency based on Cycle-GAN\cite{zhu_unpaired_2017} in order to achieve conversion of jazz, classical and pop music, while using two discriminators to determine whether the generated music belongs to the merged set of two music domains. CycleBEGAN\cite{wu_singing_2018} uses BEGAN network\cite{berthelot_began_2017} to stabilize the training process, while introducing jump connections to improve the clarity of melody and lyrics, and recursive layers to improve the accuracy of pitch to achieve male and female voice conversion. In addition, Jin et al.\cite{jin_style-specific_2020} used the LSTM network as a generator, adding music rules as a reward function in reinforcement learning to a control network that takes advantage of the AC (Actor Critic) reinforcement learning algorithm to select for diversity of elements and avoid the drawback of GAN networks generating samples that are too similar, thus improving the creativity of stylistic music generation. The same goes for Play as You Like\cite{lu_play_2019} which implements the conversion of classical, jazz and pop music. For audio generation, WaveGAN\cite{donahue_adversarial_2019} achieved the first attempt to apply GAN to the synthesis of raw waveform audio signals, and GANSynth\cite{engel_guularjani_2019} improved the method by generating the entire sequence in parallel rather than sequentially, which appended a one-hot vector representing pitch and achieved independent control of pitch and timbre based on a PGGAN networ\cite{han_learning_2019}, with the final results 50,000 times faster than the standard Wavenet\cite{oord_wavenet_2016}, and the generated music clips are better than Wavenet and WaveGAN\cite{donahue_adversarial_2019}.

Despite the outstanding performance in music generation, GAN still have shortcomings:

\begin{enumerate}
\item [(1)]They are difficult to train
\item [(2)]They are poorly interpretable, and there is still a lack of high-quality research results on how GANs can model text-like data or musical scores.
\end{enumerate}
\subsubsection{Variational Auto-Encoder}\label{subsubsec6}
\begin{figure}[h]%
\centering
\includegraphics[width=0.45\textwidth]{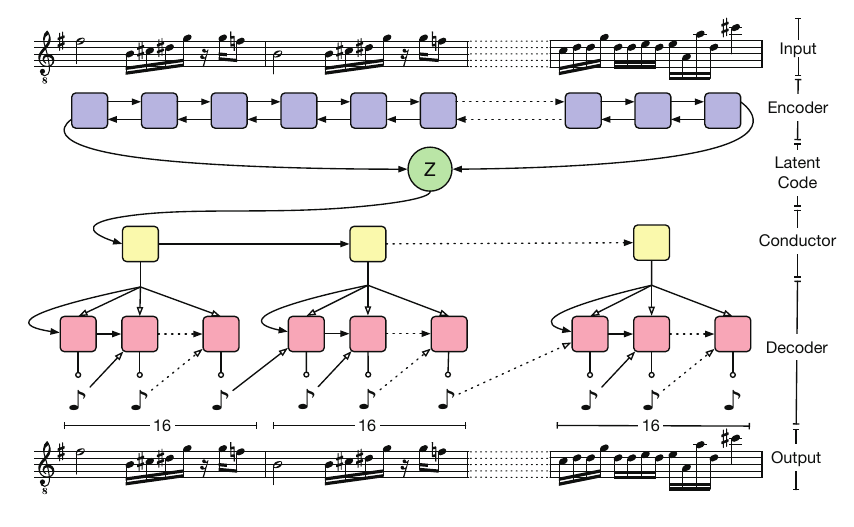}
\caption{\textcolor{black}{Schematic of the hierarchical recurrent Variational Autoencoder
model, MusicVAE\cite{roberts_hierarchical_2018}.}}\label{fig_vae}
\end{figure}
Variational Auto-Encoder (VEA) is essentially a compression algorithm for encoders and decoders, which has been able to analyse and generate information such as pitch dynamics and instrumentation in polyphonic music. The use of VAE aims to solve two problems\cite{rivero_classical_2020}: music recombination and music prediction, but when the data is multimodal, VAE does not provide a clear mechanism for generating music, usually using a hybrid coding model based on VAE, such as a combination of VAE networks and RNN networks.

Some typical examples are MIDI-VAE\cite{brunner_midi-vae_2018}, Figure \ref{fig_vae} and MusicVAE\cite{roberts_hierarchical_2018}. Among them, MIDI-VAE constructs three pairs of coders/decoders sharing a potential space to automatically reconstruct the pitch, intensity and instrumentation of a musical composition for musical style conversion. This work represents the first successful attempt to apply style conversion to a complete musical composition, but the small amount of data used for style feature training results in short lengths of the generated music. MusicVAE uses hierarchical decoder to improve the modelling of sequences with long-term structure, using a bidirectional RNN as an encoder, allowing the model to interpolate and reconstruct well. Then, GrooVAE\cite{gillick_learning_2019} as a variant of MusicVAE, generates drum melodies by training on a corpus of electronic drums recorded from live performances. \textcolor{black}{Wei et al.\cite{wei_learning_2022} did not adopt an end-to-end approach to learning hierarchical representations but instead introduced a novel model based on EC$^{2}$ -VAE. This model utilizes a context-constrained learning approach for long-term symbolic music representations, employing contrastive learning methods and a hierarchical prediction model to train and optimize these long-term representations. It can decompose the pitch and rhythm components of melodies (32 bars) without increasing the dimensions of the long-term representations. Dubnov et al.\cite{dubnov_deep_nodate} designed a vanilla polyphonic VAE using only linear layers to learn a latent representation of the musical surface. They presented a theoretical framework of musical creativity and surprisal formulated in terms of information theoretical relations between fullrate (high-fidelity) encoding of the musical data, and a lower complexity latent encoding that models reduced informational aspects of musical structure.}

For the generation of Eastern popular and folk music, MG-VAE\cite{luo_mg-vae_2019} was the first study to apply deep generative models and adversarial training to oriental music generation, which was able to convert music into folk music with a specific regional style. Jiang et al.\cite{jiang_stylistic_2019} developed a music generation model with style control. This method is based on MusicVAE and utilizes global and local encoders to construct three music generation models. It attempts to find an appropriate approach to represent musical structure at the levels of bars, phrases, and songs. Ultimately, it can generate melodies with style control.

MahlerNet\cite{lousseief_mahlernet_2019} constructed a conditional VAE to model the distribution of latent states. Two bidirectional RNN networks (one considering music reconstruction and one considering context) constitute an encoder, and the decoder outputs duration, pitch, and instrument, realized the use of any number of instruments to simulate polyphonic music of any length. MIDI-Sandwich2\cite{liang_midi-sandwich2_2019} introduces a hierarchical multimodal fusion generative VAE (MFG-VAE) network based on RNN. Initially, it utilizes multiple independent Binary VAE (BVAE) models with non-shared weights to extract feature information for different instruments in various tracks. MGF-VEA, serving as the top-level model, blends features from different modes, and finally, multi-track music is generated through the decoding of BVAE. This method improves the refinement method of BinaryMuseGAN\cite{dong_convolutional_2018}, transforms the binary input problem into a multi-label classification problem, and solves the problem of the difficulty of distinguishing the refining steps of the original scheme and the difficulty of descending the gradient. MuseAE\cite{valenti_learning_2020} applied Adversarial Autoencoders to music generation for the first time. \textcolor{black}{The advantage of this approach over VAE lies in its use of adversarial regularization instead of KL divergence, which, in principle, can enforce it to adhere to any probability distribution, thereby offering greater flexibility in selecting the prior distribution for latent variables.} The Jukebox\cite{dhariwal_jukebox_2020} built a system that can generate a variety of high-fidelity music in the original audio domain, but this method still requires 9 hours to render one minute of music.
\subsubsection{Transformer}\label{subsubsec7}
A piece of music often has multiple themes, phrases or repetitions on different time scales. Therefore, generating long-term music with complex structures has always been a significant challenge. The music generation system is usually based on a sequential RNN (or LSTM, GRU, etc.) network, but this mechanism brings about two problems: Calculating from left to right or calculating from right to left limits the parallel ability of the model; Part of the information will be lost during the sequence calculation. This led Google to propose the classical network architecture Transformer\cite{vaswani_attention_2017}, which uses an attention mechanism instead of the traditional Encoder-Decoder framework that combine the inherent patterns of CNNs or RNNs. The core idea of this architecture is to use the Attention mechanism to indicate correlations between inputs, with the following advantages:

\begin{enumerate}
\item [(1)]Data parallel computing
\item [(2)]Visualization of self-reference
\item [(3)]Solve the problem of long-term dependence better than RNN, and has achieved great success in the task of maintaining long continuity in timing.
\end{enumerate}

\begin{figure}[h]%
\centering
\includegraphics[width=0.5\textwidth]{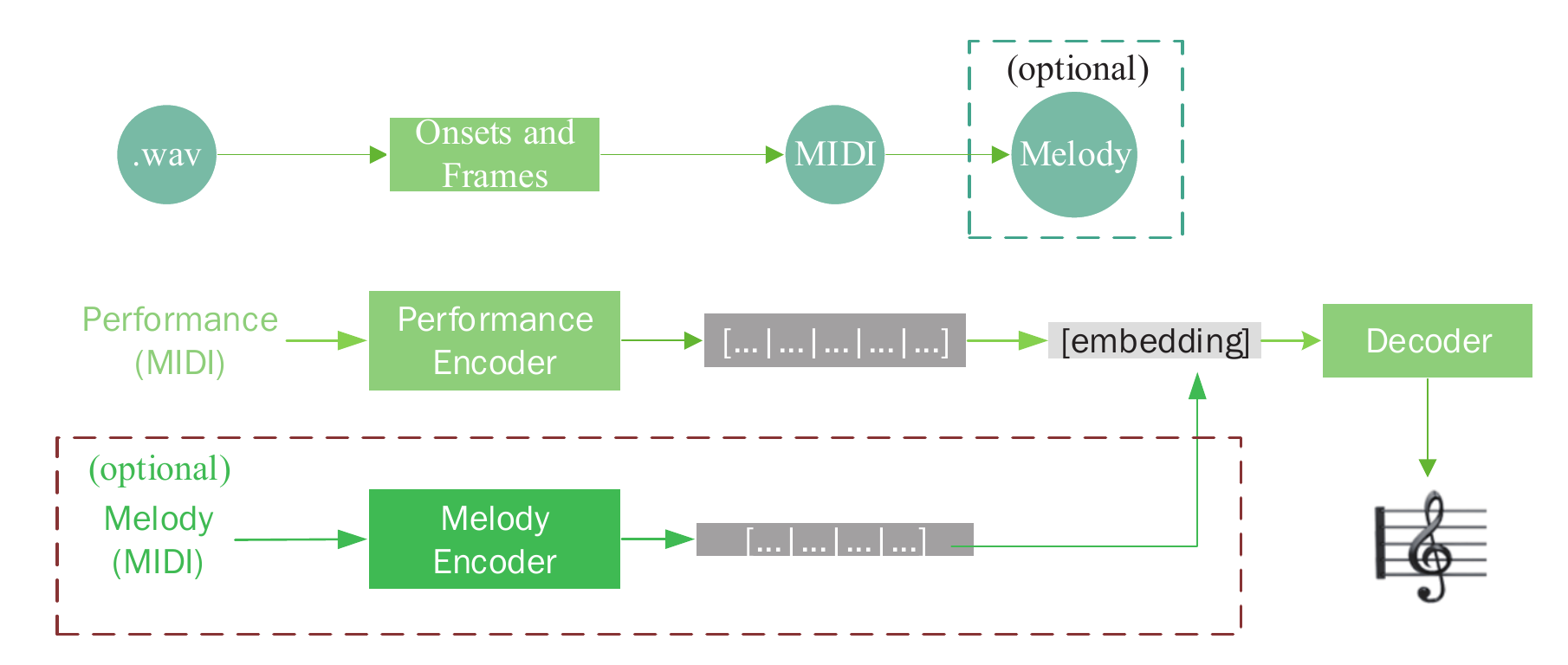}
\caption {The flow chart of Transformer self-encode: First transcribes the .wav data file into a MIDI file using the Onsets and Frames framework, which is then encoded into a performance representation to be used as input. The output of the performance encoder is then aggregated across time and (selectively) combined with melodic embedding to produce a representation of the entire performance, which is then used by the Transformer decoder when reasoning\cite{Musenet}.}\label{fig_transformer}
\end{figure}

However, the feasibility of this model in music generation tasks is limited because the spatial complexity of the intermediate vectors representing relative positions in the algorithm is extremely high (quadratic in the sequence length). \textcolor{black}{Music Transformer\cite{huang_music_2018} significantly reduces the spatial complexity of the intermediate vectors representing relative positions to the order of sequence length, making it applicable for piano music composition. However, for longer music generation, the resulting compositions suffer from poor musical quality and structural disorder.} LakhNES\cite{donahue_lakhnes_2019} used transfer learning to generate chip music by pre-training in a large-scale dataset using Transformer's latest extension Transformer-XL, and then fine-tuning in a small-scale chip music dataset. MuseNet\cite{Musenet} used the same network as GPT-2 (a large Transformer-based language model\cite{budzianowski_hello_2019}), which is trained using the Transformer's recomputed and optimised kernel to generate a four-minute musical composition consisting of ten different instruments, but the model does not necessarily arrange the music according to the input instruments, generating each note by calculating the probability of all possible notes and instruments. Figure \ref{fig_transformer} shows the flow chart of this Transformer self-encoder. 

In addition, some researchers have used Transformer in combination with other models to address problems such as the poor musicality of long sequences. Guan et al.\cite{guan_gan_2019} proposed a combination of Transformer and GAN, first using a self-attentive mechanism based on GAN to extract spatial and temporal features of music to generate consistent and natural monophonic music, and then using a generative adversarial structure with multiple branches to generate multi-instrumental music with harmonic structure between all tracks. Zhang et al.\cite{zhang_learning_2020} also used the Transformer in combination with GAN to make the generation of long sequences guided by the adversarial goal, which provides powerful regularisation conditions that can force the Transformer to focus on learning the global and local structure of the music. Transformer VAE\cite{jiang_transformer_2020} combines Transformer with VAE, effectively addressing the limitations of VAE in handling time series structures and the non-explanatory nature of Transformer's latent states. Choi et al.\cite{choi_encoding_2020} similarly used the Transformer (decoder) to obtain a global representation of the music, thus better control over aspects such as performance style and melody.

\textcolor{black}{Several recent works have introduced innovative music generation frameworks, including user-friendly interfaces, Transformer-based methodologies, and novel representations, which have demonstrated improvements in music style fidelity, enriched melodic content, and faster sequence generation. Guo et al.\cite{guo_musiac_2022} introduce a new music generation framework with a user-friendly interface for infilling, utilizing a transformer-based approach that supports customizable control tokens, including tonal tension and track polyphony. Their research explores the impact of these tokens on music generation, demonstrating improved stylistic fidelity compared to previous methods. The model is made available in a Google Colab notebook for interactive music creation. Zou et al.\cite{zou_melons_2021} proposed  a transformer based framework MELONS that can generate enjoyable long-term melodies with obvious structure and rich contents by using  a structure generation net and a melody generation net.  One of the key ideas is to represent structure information using graph which consists of eight hierarchical relations among pairwise bars. Dong et al.\cite{dong_multitrack_2023} propose a new multitrack music representation that encoded each musical event as a tuple of six variables (i.e., type, beat, position, pitch, duration, and instrument) to significantly shorten the sequence length of multitrack music and also allow a diverse set of instruments. Furthermore, a decoder-only transformer model with multi-dimensional input and output spaces called Multitrack Music Transformer (MMT) was proposed to generate longer sequences at a faster inference speed. Besides, Yu et al.\cite{yu_museformer_2022} proposed Museformer with a novel fine- and coarse-grained attention. The fine-grained attention is applied to the structure-related bars for learning the structure-related correlations, and the coarse-grained attention is applied to the summarization of the other bars for getting a sketch of them. }

\subsubsection{Evolutionary Algorithms}\label{subsubsec8}
The field of evolutionary computing encompasses various techniques and methods inspired by natural evolution. At its core lies the Darwinian search algorithm based on highly abstracted biological models. The advantage of applying them to music generation systems is that they can be optimized in the direction of high relative quality or adaptability, thus compressing or extending the data and processes required for the search. It provides an innovative and natural means of generating musical ideas from a set of specifiable original components and processes\cite{miranda2007evolutionary}.  EA requires three basic criteria to be considered in advance: 1. Problem domain: defining the type of music to be created; 2. Individual Representation: the representation of the music, melody, harmony, pitch, duration, etc.; 3.Fitness Measure.
\begin{figure}[h]%
\centering
\includegraphics[width=0.45\textwidth]{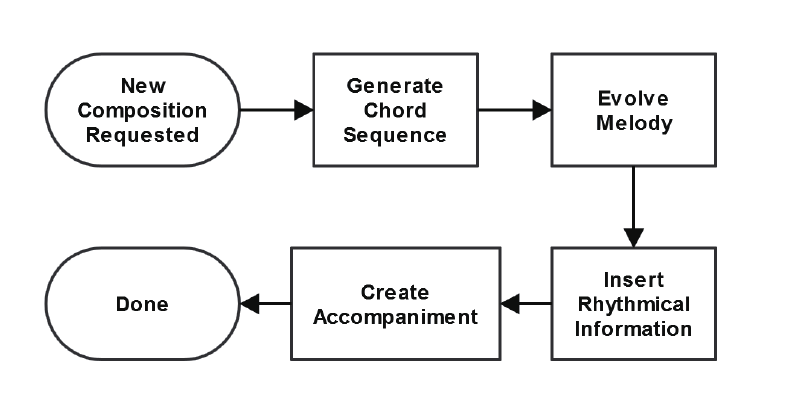}
\caption{\textcolor{black}{Steps for generating a composition using evolutionary algorithms\cite{scirea_metacompose_2016}.}}\label{fig_ea}
\end{figure}
Fitness measure is a very important part of genetic algorithms to evaluate the goodness of a solution, where the fitness statistics of different music generation systems are shown in Tab \ref{tab2}. In music generation systems, fitness measure includes interactive adaptation, similarity to a music corpus, rule-based metrics, etc. Early evolutionary algorithms commonly used interactive adaptation measure, which requires significant human intervention and can suffer from adaptation bottlenecks; poor robustness and wider applicability in music corpus or rule-based fitness measures, as well as limitations on the creative potential of the system.

For melody generation, Munoz et al.\cite{munoz_memetic_2014} employ unfigured bass harmonization techniques to create bass melodies. They use harmonic rules as an adaptability metric, taking the bass line as input, and in a parallel manner, they find subspaces for suitable melodies. By utilizing a local composer intelligent agent with self-adaptation, they produce high-quality four-part musical compositions. Jeong et al.\cite{jeong_multi-objective_2017} proposed a multi-objective evolutionary algorithm for automatic melodic synthesis that produces multiple melodies at once, which measures the psychoacoustic effects of the resulting music based on the stability and tension of the sound as an adaptive metric. Lopes et al.\cite{lopes_combining_2017} proposed an automatic melodic synthesis method based on two adaptive functions, one defined using Zipf's law\cite{zipf_human_2016} and the other using Fux's law\cite{fux_study_1965}. GGA-MG\cite{farzaneh_gga-mg_2020} uses LSTM as an adaptive function, first providing a sufficient dataset of random and artificial samples using an initial Generative Genetic Algorithm (GGA), and then using a Bi-LSTM neural network for training as an adaptive function for the main GGA to generate melodies similar to those of human compositions.

In arranging music, MetaCompose\cite{scirea_metacompose_2016}, figure \ref{fig_ea}, was used to generate real-time music, using a multi-objective adaptive function to construct a chord generator, melody generator, and accompaniment generator, but the method lacked the generation of harmonies. \textcolor{black} {EvoComposer\cite{de_prisco_evocomposer_2020} EvoComposer [81] enhances this approach by generating harmonies for four voice types (tenor, bass, soprano, alto) based on a given choice. It defines an adaptability function that adheres to classical music rules while incorporating statistical information extracted from an existing music corpus to absorb specific composer styles.} Olseng\cite{olseng_co-evolving_2018} used four adaptation measures: local, global melodic features and local, global harmonic features with the aim of making the melodies generated by the algorithm sound harmonious and pleasant. Kaliakatsos\cite{kaliakatsos-papakostas_interactive_2016} uses Particle swarm optimization (PSO) to converge on user-preferred rhythmic and pitch features, and then uses a genetic algorithm to compose music based on these features in a 'top-level' evolution.

\section{Evaluation}\label{sec5}
In the field of music generation, evaluating the effectiveness of generation is a major challenge. Existing evaluation methods mainly combine subjective and objective evaluation, where subjective evaluation includes Turing tests, questionnaires, etc., and objective evaluation includes quantitative indicators based on musical rules, similarity to corpus or style, etc. Agres et al.\cite{manzelli_end_2018} provide an overview of current evaluation methods for music generation systems, providing provides motivation and tools. These metrics are interrelated and together influence the construction of the music generation system and the quality of the resulting pieces.
\subsection{Subjective evaluation}\label{subsec5.1}
Subjective evaluation of music generation systems is usually done by performing human listening tests to evaluate the performance of the system in terms of melodic output. This method can be used in a similar way to the Turing test or questionnaires. For example, in MG-VAE\cite{luo_mg-vae_2019}, listeners are guided to rate the musicality (i.e. whether the song has a clear musical style or structure) and stylistic significance (i.e. whether the style of the song matches a given regional label) of the generated music, combined with several quantitative metrics based on musical rules to evaluate the merits of the music generation system. Chen et al.\cite{chen_effect_2019} define a manual evaluation metric that includes Interactivity i.e. whether the interaction between chords and melody is good, complexity i.e. whether the piece is complex enough to express the theme, and structure i.e. whether there is some repetition, forward movement, variation in the sample. Performance RNN collects people's ratings of the generated music; \textcolor{black}{Jaques et al.\cite{jaques_tuning_2017, wang_performancenet_2019, liu_lead_2018} combine human ratings to judge the performance of the music generation system. In \cite{jaques_tuning_2017}, a user study is conducted via Amazon Mechanical Turk in which participants are asked to rate which of two randomly selected melodies they preferred on a Likert scale. In \cite{wang_performancenet_2019}, an online user study is conducted to evaluate the result of arrangement generation. After listening to the music, respondents are asked to vote for the best model among the three, in terms of harmonicity, rhythmicity and overall feeling, respectively. Moreover, they are asked to rate each sample according to the same three aspects. In \cite{liu_lead_2018}, they evaluate the generated music clips by a user study and compare the mean opinion score (MOS).  Specifically, 156 adults of which 41 are music professionals or students major in music are recruited and asked to rate the music pieces following metrics in a five-point Likert scale.}

While human feedback may be the most reasonable method for evaluating generated music, it has certain limitations. Human users are unable to rate a large volume of music, and user fatigue may occur within the first few minutes of rating or selecting music. Additionally, due to individual differences in audience taste, the uncertainty in music ratings or selections increases, resulting in inconsistent ratings. While Sturm et al.\cite{sturm_machine_2019} extensively discuss the advantages of evaluating music in the form of concerts or by extending concerts to music competitions, such as setting a concert as one of the most natural ways to experience music, which can to some extent reduce the fatigue of subjects compared to laboratory settings, this method inevitably introduces differences due to variations in performers, venues, and environments.

\subsection{Objective evaluation}\label{subsec5.2}
Objective evaluation of music generation systems includes similarity to a corpus or style, quantitative indicators based on musical rules, etc. Similarity is central to the success metrics in music generation systems, therefore, an important challenge becomes finding the right balance between similarity and novelty or creativity. Evaluating the creativity (sometimes equated with novelty) of generated music is a complex topic and is discussed at length by Agres et al.\cite{agres_evaluation_2016}

Different objective evaluation criteria can be chosen depending on the type of generated music. Zhang et al.\cite{zhang_learning_2020} devised seven objective metrics to compare generated musical compositions with real compositions, these include: number of different note repetitions, count of different pitches, number of triplets, etc. GANSynth\cite{engel_guularjani_2019} compared generated music with real compositions by using manual evaluation, counting the number of difference Bins (a measure of generated music fragment diversity), initial scores (to evaluate the GAN network), pitch accuracy, pitch entropy, etc. JazzGAN\cite{trieu_jazzgan_2018} employs three metrics, including the pattern collapse metric for observing intervals and durations of repeated and incoherent notes, the creativity metric to measure the extent to which generated segments replicate sequences from the corpus and the number of variant sequences, and the harmony and chordal metric to evaluate the relationship between the generated musical sequence and a given chord sequence in order to assess the generative system. MuseGAN\cite{dong_musegan_2018} has designed two aspects for evaluating the differences between generated music and real music in a single track and between different tracks, which include within a track: the proportion of empty bars, the number of pitch classes per bar, the proportion of eligible notes, and the ratio of notes per 8-beat or 16-beat interval, and between tracks: the pitch distance between tracks. \textcolor{black}{In \cite{jaques_tuning_2017}, objective metrics are selected to assess the results, including 1) the number of notes belonging to some excessively repeated segment; 2) the ability to play in key, resolve melodic leaps, and play motifs; 3) the number of melodies that start with the tonic note; 4) melody auto-correlation, and 5) repeated motifs. In \cite{wang_performancenet_2019}, four objective metrics are selected: 1) Empty bars; 2) used pitch classes; 3 )qualified note and 4) tonal distance.
}

A major challenge in the objective evaluation of music production systems is that there is no single objective evaluation criterion for judging the 'goodness' of a system. These criteria are limited in that they are only applicable to the current type of music generation task and cannot be applied to all music generation systems, and their reasonableness has yet to be considered.

\section{Analysis \& Perspectives}\label{sec6}
\subsection{Current state of intelligent music generation}\label{subsec6.1}
Due to the fact that music generation systems (1) can provide a far greater volume of content than composers can offer, their cost per minute of music is much lower than that of composers; (2) they can customize and personalize music based on user settings; (3) they can inspire composers and provide assistance; (4) they can be used for music education, among other applications. Therefore, this field has very extensive prospects for development and commercial value. The company cases that have already landed include Melodrive's focus on instant music generation in game scenarios, Amper Music to generate personalized customized original music in a few seconds, and Ambient Generative Music to automatically generate personalized music in different environments.

With the development of artificial intelligence, there are also certain difficulties in the rapid development of artificial intelligence composition. The current music generation systems can generate polyphonic music with a higher quality and a shorter period of time containing different styles and instruments in a shorter period of time\cite{jin_style-specific_2020}. However, the construction of a long time series of music fragments with a complete structure\cite{jiang_transformer_2020} and the evaluation of generated music\cite{agres_evaluation_2016} have become a common problem for artificial intelligence composition researchers. Although the latest music generation system can generate longer-term music fragments, such as MuseNet, which can use 10 different instruments to generate 4-minute music works. \textcolor{black}{However, the latest music generation algorithms still struggle to effectively handle time series issues. As the generated music segments become longer, the musicality of the produced segments deteriorates, resulting in increasingly chaotic structures or higher levels of repetition.} Currently, researchers are focusing on optimizing generation algorithms or imposing constraints on generated music segments based on music theory knowledge to alleviate the temporal dependency issues in music generation systems.

\subsection{Comparative Analysis of Music Generation Research in Eastern and Western Contexts}\label{subsec6.2}
From a technological perspective, there is a noticeable disparity in the development of music generation between the Western and Eastern context. \textcolor{black}{While western music generation research has reached a relatively mature stage, with companies like OpenAI\cite{Musenet}, Google\cite{Chrome Music Lab}, and AIVA\cite{AIVA} offering publicly accessible online music generation tools, research in the Eastern world is still in its nascent stages in this field. Although some Eastern enterprises and research institutions in China, Japan and India are actively exploring this field\cite{zhu_xiaoice_2018, choi_chord_2021, lee_polyphonic_2018, mangal_lstm_2019, shin_melody_2017, wada_sequential_2018}, research on music generation with a focus on Eastern indigenous music genres remains relatively limited. This discrepancy can be attributed to several factors, represented by differences in musical structure and thought development due to cultural backgrounds, the scarcity of Eastern music databases, the absence of effective evaluation methods , and the lack of interdisciplinary professionals bridging the gap\cite{jie_compare}.}

\textcolor{black}{Cultural differences play a profound role in shaping the technological characteristics of symbolic music generation. Eastern music differs notably from Western classical music (especially symphonic compositions) in terms of structure, musical thinking, and artistic expression \cite{hu_across, son2015pagh, mok2014east}. Regarding musical structural power and thinking, Eastern music genres exhibits a distinct understanding that typically construct single-line monophonic music and emphasize the aesthetics of the music. Eastern classical music. For example, Chinese classical music primarily revolves around pentatonic scales and there is a preference for ensemble playing with a group of vocal parts or instruments in traditional Japanese music\cite{matsue2015focus}. Conversely, Western music tends to be polyphonic with an emphasis on major and minor scales, focusing on harmonic logic and using techniques like repetition, contrast, and extension to create atmospheres. Moreover, in the context of artistic expression, Eastern and Western music genres vary significantly. Western music pay more attention to profundity and seriousness, highlighting the contrast between subjectivity and objectivity. In contrast, Eastern traditional music places more significance on the unity of subjectivity and objectivity\cite{veblen_community}. And there exists considerable content for improvised performance\cite{nooshin2006improvisation}, resulting in more flexible and diverse rhythm. From an information theory perspective, this increases the difficulty of labeling and analyzing traditional Eastern music data. Considering the intrinsic cultural differences like structural, content, and performance differences between Eastern and Western music, music generation systems need extensive adjustments and optimization to be applicable to either, thereby reducing the universality between Eastern and Western music generation systems.}

The lack of diverse and high-quality Eastern music corpora is another significant factor restricting the development of music generation systems primarily focused on Eastern music genres. While datasets like MG-VAE\cite{luo_mg-vae_2019} have gathered folk music from various regions in China and collections like XiaoIce Band\cite{zhu_xiaoice_2018} and CH818\cite{hu_mood_2017}  cover Chinese pop music, these data resources are still relatively limited. Compared to a variety of predominantly Western music datasets, the insufficiency in both the quantity and quality of Eastern corpora hinders music generation algorithms and models from fully grasping the unique features of Eastern music, thereby impacting the quality of generated music segments. \textcolor{black}{In other Eastern countries, such as India, despite the presence of a rich traditional music culture, there are still challenges in systematically organizing and digitizing it\cite{srinivasamurthy2021saraga}. Similarly, music in the Middle Eastern region, such as Arabic and Persian music, boasts a profound history and unique characteristics, but is limited by challenges related to digitization and data integration\cite{repetto2018open}. Data pertaining to these musical forms remains relatively scarce. Given that the music of these regions features distinct scales, rhythms, and instrument characteristics, the lack of comprehensive music databases hampers the development of generative technologies tailored to these musical styles.} Furthermore, due to variations in development levels and stages, the intricate and complex task of collecting various types of Eastern ethnic music across different cultures and regions is encountered. And issues related to song copyrights also pose constraints on the advancement of music generation research.

Another challenge is the lack of effective evaluation methods and standardized objective assessment criteria for music generation systems. Firstly, this implies that, in automatically generated music segments, there is often a reliance on human ratings to identify those with superior structure and sound quality. \textcolor{black}{Meanwhile, eastern music encompasses a diverse and unique musical tradition, including various genres and styles from China, India, Japan, and other\cite{howard_music_2016}.} Secondly, in comparison to the extensive research on Western music genres and structures, the analysis of various Eastern music genres remains limited, making it difficult to obtain objective data suitable for evaluating the effectiveness of Eastern music generation systems. These two issues collectively constrain the further advancement of Eastern music generation research, as the absence of effective evaluation methods and limited structural analysis hinder the improvement of the quality of Eastern music generation systems.

\textcolor{black}{It is well-established that music theory and strict rules potentially play a crucial role in enhancing generative systems\cite{carnovalini_computational_2020}.} However, in practice, most researchers in the East studies tend to specialize in a singular domain, lacking interdisciplinary knowledge to effectively integrate the foundational principles of music with cutting-edge generative models. \textcolor{black}{Such imbalance has resulted in unequal progress in the subfields of Eastern music generation and most researchers channel their primary efforts towards optimizing generative algorithms, inadvertently neglecting the potential impact of music theory in refining the generative systems. Moreover, due to the intricacies of music theory, it proves exceptionally challenging for researchers to independently undertake the comprehensive learning and adept modeling of musical theoretical constructs without professional guidance.}

\textcolor{black}{In summary, research on music generation with a focus on Eastern indigenous music genres faces inherent challenges such as cultural disparities, insufficient musical databases, inadequate evaluation criteria, and a lack of interdisciplinary knowledge. In order to address these challenges, it is imperative to augment research funding and educational programs, raise public awareness. Concurrently, there is a necessity to bolster interdisciplinary collaboration, curate data comprehensively, formulate standardized evaluation criteria, and promote international cooperation. These measures are essential for advancing the research and development of high-quality music generation systems with cultural relevance. Furthermore, it is worth noting that in today's globalized society, musical cultures have been intermingling and fusing with each other. As music evolves, there is a noticeable shift in perspective from the binary distinction between the East and the West towards differentiating and contrasting music from the standpoint of local, traditional, and modern characteristics. This trend serves to emphasize the diversity and complexity of music, and it also facilitates the further development of music generation models. Therefore, it is anticipated that future research will engage in in-depth discussions and analyses from this perspective.}

\subsection{Trends in Artificial Intelligence Music}\label{subsec6.3}
\textcolor{black}{Whether it's addressing the limitations of Eastern music or exploring the commonalities between Eastern and Western music generation,} both can serve as directions for the future development and enhancement of music generation systems. The ultimate goal is to generate beautiful songs that meet individual preferences and possess complete structures. By examining the development and current state of AI composition, the trends in the evolution of music generation systems are as follows.

\subsubsection{Music Generation Technology Moves towards Maturity}\label{subsubsec6.3.1}
Currently, music generation systems predominantly focus on generating either lyrics or melodies. For cross-modal music generation, such as simultaneous generation of music and lyrics or music and video, this will be a crucial direction in the future development of artificial intelligence in music. \textcolor{black}{This direction necessitates the extraction and fusion of features across different modalities, presenting certain technical challenges.} In the future development of music generation systems, improvements can be standardized through the use of more robust music data architectures, including Piano-roll, MIDI events, sheet music, and audio, along with more standardized testing datasets and evaluation metrics.

Furthermore, given that deep learning architectures are the mainstream generation methods for music generation systems, enhancing the interpretability of deep networks and their controllability becomes one of the future directions for music generation technology. Additionally, addressing the issue of temporal dependencies in generating music, focusing on segment-level audio generation rather than note-level audio generation, represents a research trend in future music generation systems.

\subsubsection{Emotional expression of music generation and its coordinated development with robot performance}\label{subsubsec6.3.2}
Music is one of the most important forms of human emotion expression. Musical emotions are conceptually regarded as an expression of human emotion that is difficult to quantify, and it undergoes rich changes with the progress of music. The current degree of music generation intelligence is at a low level, based more on the perspective of signal analysis, without introducing a human recognition system for musical emotions, without anthropomorphic musical composition thinking, and with human-computer interaction systems limited to shallow information exchange. Therefore, the integration of multi-channel information such as computer vision and computer hearing to recognise the human spectrum of musical emotions and audio expression systems, and then the optimisation of the emotional dimension of music generation using techniques such as deep learning is the main goal of the construction of future interactive intelligent composition systems.

Furthermore, research in the field of music visualization based on robotic systems\cite{lin_towards_2018} and emotionally interactive music therapy robots\cite{tapus_role_2009} integrates robotics technology with musical expression, with robots as the central element. In the future, achieving intelligent composition and collaborative performance of music robots under the context of affective computing is a significant highlight in the integrated development of intelligent music generation and performance.

\section{Conclusion}\label{sec7}
The involvement of artificial intelligence in artistic practices has become a common phenomenon, and using AI algorithms to generate music is a highly active research area.

\textcolor{black}{This paper systematically outlines the latest developments in the field of music generation from the perspective of algorithm types and briefly introduces various forms of musical representation and evaluation criteria.  Furthermore, the paper analyzes the current state of AI composition on a global scale, particularly in the contexts of  Eastern and Western traditional music, and compares the distinct developmental features of the two. In particular, this paper examines the opportunities and challenges for future music generation systems. It is hoped that this review contributes to a better understanding of the current status and trends in AI-based music generation systems.}
\section*{Declarations}

% Some journals require declarations to be submitted in a standardised format. Please check the Instructions for Authors of the journal to which you are submitting to see if you need to complete this section. If yes, your manuscript must contain the following sections under the heading `Declarations':

\begin{itemize}
\item Competing Interests
The authors declare that they have no conflict of interest.
\item Ethics approval 
This article does not contain any studies with human participants or animals performed by any of the authors.
\item Availability of data and materials
Not applicable
\item Code availability
Not applicable
\item Authors' contributions
Methodology: Lei Wang; Formal analysis and investigation: Ziyi Zhao; Writing - original draft preparation: Ziyi Zhao, Hanwei Liu; Data curation: Junwei Pang; Writing - review and editing: Yi Qin, Song Li, Ziyi Zhao; Supervision: Qidi Wu, Lei Wang Funding acquisition: Lei 
\item Acknowledgment
\textcolor{black}{Figure 4-11 are cited from relevant paper on music generation, and we extend our sincere appreciation to the authors for their contribution.}
\end{itemize} 
\noindent

%%===========================================================================================%%
%% If you are submitting to one of the Nature Portfolio journals, using the eJP submission   %%
%% system, please include the references within the manuscript file itself. You may do this  %%
%% by copying the reference list from your .bbl file, paste it into the main manuscript .tex %%
%% file, and delete the associated \verb+\bibliography+ commands.                            %%
%%===========================================================================================%%
% \bibliography{sn-bibliography}% common bib file
% \bibliography{bib20220105}% common bib file
\bibliographystyle{unsrt}
%% if required, the content of .bbl file can be included here once bbl is generated
% \input generate_bbl.bbl

%% Default %%

\end{document}